\documentstyle[preprint,aps,epsf]{revtex}
\begin{document}
\draft
\preprint{\setlength{\baselineskip}{2.6ex}\hfil
\vbox{\hbox{TRI--PP--97--13}
\hbox{hep-ph/9705293}
\hbox{July 30 1997}}}

\title{Polarization Measurements and $T$-Violation
in Exclusive Semileptonic $B$ Decays}

\author{Guo-Hong Wu$^a$\footnote{\tt{gwu@alph02.triumf.ca}
\vspace{-.15in}}, Ken Kiers$^b$\footnote{\tt{kiers@bnl.gov}
\vspace{-.15in}} and
John N. Ng$^{a}$\footnote{\tt{misery@triumf.ca}}}

\address{$^a$TRIUMF Theory Group\\4004 Wesbrook Mall, Vancouver, B.C.,
V6T 2A3 Canada\\ {\rm and} \\$^b$Department of Physics
\\Brookhaven National Laboratory, Upton, NY 11973-5000, USA}
\maketitle

\begin{abstract}
We provide a general analysis of time reversal invariance violation in the
exclusive semileptonic  $B$ decays $B \to D \ell \overline{\nu}$ and
$B \to D^* \ell \overline{\nu}$.  Measurements of the lepton and $D^*$
polarizations can be used to search for  and identify non-standard model
sources of $T$ violation.  Upper limits are placed on the $T$-odd
polarization observables in both the supersymmetric $R$-parity conserving
and $R$-parity breaking theories, as well as in some non-supersymmetric
extensions of the standard model, including multi-Higgs-doublet models,
leptoquark models, and left-right symmetric models.  It is noted that many
of these models allow for large $T$-violating polarization effects which
could be within the reach of the planned $B$ factories.
\end{abstract}

\newpage

\section{Introduction}

The origin of $C\!P$ violation remains one of the mysteries
of elementary particle physics today,
although the observed $C\!P$-violating phenomena in the kaon system
are consistent with the standard model Cabibbo-Kobayashi-Maskawa
(CKM)~\cite{CKM} paradigm.
One of the principal goals of the planned $B$-factories is
to test the standard model (SM) parameterization of $C\!P$ violation through
precision measurements in several of the hadronic
decay modes of the $B$ meson \cite{stone}.
Any deviation from the SM prediction would be a signal of new physics.
Such a signal would of course be welcome,
since the gauge hierarchy problem of the SM has led to a widely held belief
that the SM is actually a low-energy
approximation to some more complete theory.
A generic feature of many extensions of the SM is
the presence of new $C\!P$-violating phases.
Given the large number of $B$'s expected at the $B$-factories, it
is clearly important to examine the various $C\!P$-odd observables
in the $B$ system in order to identify those which are sensitive
to new physics.  Of particular interest are those observables
which receive negligible contributions from SM sources.

In this work we present a detailed analysis of several of the
$T$-odd\footnote{We assume $C\!PT$ invariance throughout and so will
use ``$C\!P$-odd'' and ``$T$-odd'' interchangeably.} observables which are
available in the exclusive semileptonic decays of $B$ mesons
to $D$ and $D^*$ mesons.
 Measurements of these observables would complement the studies of
$C\!P$ violation in the hadronic decay modes and could serve as valuable
tools in order to identify the Lorentz structure of any observed new effects.
In a previous paper~\cite{wkn} we have shown that one can define
$T$-odd polarization observables (TOPO's) in the decays
$B$$\to$$D^{(*)}\ell\overline{\nu}$ ($\ell$$=$$e,\mu, \tau$) which are
sensitive separately to effective
scalar, pseudoscalar, and right-handed current interactions.
In the present work we will provide a more comprehensive analysis
of these observables in addition to considering the prospects
in various models for measuring a positive signal.

It has long been known that the semileptonic decays of pseudoscalar
mesons provide an ideal place in which to search for non-SM $T$-violating
signals~\cite{sakurai}.
One of the best studied of these $T$-odd observables is the muon
transverse polarization in the decay $K^+\to\pi^0\mu^+\nu_\mu$ ($K_{\mu3}^+$),
defined by
\equation
	P_{\mu}^{\bot}\equiv \frac{d\Gamma(\vec{n})-d\Gamma(-\vec{n})}
		{d\Gamma(\vec{n})+d\Gamma(-\vec{n})},
	\label{eq:kmu3transpol}
\endequation
where $\vec{n}$ is the projection of the muon spin normal
to the decay plane.  Experiments at the Brookhaven National Laboratory
give the combined result \cite{BNL1981}
\equation
	P_{\mu}^{\bot} = (-1.85\pm 3.60)\times 10^{-3},
\endequation
which translates into an upper bound of $.9\%$ at the $95\%$ confidence
level.  Current efforts at the on-going KEK E246 experiment \cite{KEK246}
and at a recently proposed BNL experiment \cite{BNLproposal} are expected to
reduce the error on this quantity by factors of $10$ and $100$, respectively.
This optimistic experimental outlook has generated much theoretical interest
in the muon transverse polarization in both the
$K_{\mu3}^+$~\cite{zhit,cheng,kmu3,CF,wn1} and $K^+\to\mu^+\nu_\mu\gamma$
($K_{\mu2\gamma}^+$)~\cite{marciano,geng,KLO,wn2} decays.

The muon transverse polarization defined above is proportional
to $\vec{s}_{\mu}\cdot(\vec{p}_{\pi}\times\vec{p}_{\mu})$,
which is the only $T$-odd quantity available in that decay.
One can define analogous quantities for the leptons in the decays
$B$$\to$$D^{(*)}\ell\overline{\nu}$ and one finds that,
neglecting tensor effects, they are sensitive to non-SM
scalar and pseudoscalar effective interactions in the $D$
and $D^*$ cases, respectively \cite{wkn}.
The $\tau$ lepton polarization in these decays has been
studied in multi-Higgs models~\cite{atwood,garisto,grossman,kunob}
and, more recently, in $R$-parity conserving supersymmetric (SUSY)
models with large intergenerational squark mixing~\cite{wn1,wkn}.
In the latter case the effect arises at one loop.
In both types of models the transverse $\tau$ polarization can be
rather large (from 10's of percent to order unity) compared to the
the muon transverse polarization in $K_{\mu3}$ decay.
One of the reasons for these seemingly large
numbers is that the polarization effects in these models are proportional
to the lepton mass.  Choosing $\ell$$=$$\tau$ can thus give a
substantial enhancement compared to the $\ell$$=$$\mu$ case.
 From an experimental point of view this means that polarization measurements
in $B_{\tau3}$ decays can achieve the same ``new physics reach'' as
analogous measurements in $K_{\mu3}$, with far fewer events.
One recent study suggests that HERA-B
could achieve an eventual sensitivity to the
transverse $\tau$ polarization on the order of a few percent~\cite{herab},
which would then be competitive -- in terms of reach -- with that
expected in the current $K_{\mu3}$ experiments.
One could in principle also study the transverse polarization
of the electrons or muons in semileptonic $B$ decays.  Since these
lighter leptons are highly energetic, however, it is in practice very
difficult to measure their polarizations.  For this reason the
electron and muon transverse polarizations will
not be considered here, although these quantities need not be small
in some extensions of the SM.

As we have noted previously~\cite{wkn}, the semileptonic $B$ decays
have a novel feature compared to the analogous $K$ decays in that
the $B$ can decay to both pseudoscalar and vector mesons.  The polarization
vector of the $D^*$, which is odd under $T$, may thus also be used
to construct TOPO's.
 There are in fact two distinct TOPO's which may be constructed
using the $D^*$ polarization and they are both sensitive to effective
right-handed current interactions.  For $\ell$$=$$\tau$
 one of the TOPO's can also depend (to a lesser extent)
on effective pseudoscalar interactions.
Combining the lepton polarization measurements and
$D^*$ polarization measurements would thus allow one to probe separately
the different Lorentz structures of non-SM sources of $T$ violation.

The outline of this paper is as follows.  In Sec.~\ref{sec:general}
 we provide a model-independent analysis of the lepton and $D^*$
polarization based on an effective lagrangian approach.
The $T$-odd $D^*$ polarization observables can be related to $T$-odd
triple-momentum correlations \cite{GV,KSW} in the four-body final state
of the decay $B \rightarrow D^*(D\pi) \ell \overline{\nu}$.
This connection is made explicit in Appendix~\ref{sec:appendb}.
In Sec.~\ref{sec:models} the maximal sizes of these
$T$-odd polarization observables are estimated in several classes of models.
In $R$-parity conserving SUSY, $T$ violation occurs at the loop level
and its effect is negligible in the absence of squark family mixings.
We demonstrate that large enhancements can occur in the
presence of squark generational mixings, giving rise
to observable $T$-odd polarization effects while escaping the
flavor changing neutral current (FCNC) bounds.
We also consider $R$-parity-violating SUSY models.  In this case
the present data place stringent limits on these TOPO's.
We then consider several non-SUSY models, giving estimates for the
maximal sizes of the TOPO's in multi-Higgs models,
leptoquark models, and left-right symmetric models.
We conclude in Sec.~\ref{sec:conclusion} with a brief discussion
and a summary of our results.

\section{General Analysis}
\label{sec:general}

In this section we provide a general
analysis of the $T$-odd polarization
observables available in semileptonic $B$ decays.
The effects of new physics
may be conveniently parameterized by an effective lagrangian
written in terms of the SM fields.
For definiteness, we will always consider the decays
$B^{-}$$\to$$D^{(*)0}\ell^-\overline{\nu}$, with $l$$=$$e,\mu,\tau$.
The analogous TOPO's for the charge conjugates of these decays
may always be obtained simply by changing the sign \cite{OK}.  One could in
principle also consider the decays of neutral $B$'s.  In these decays
the electromagnetic final state interactions (FSI's) could mimic the
$T$-odd observables which we will be studying.  This effect is, however,
small on the scale of the experimental
sensitivity expected at the upcoming experiments and could probably be
ignored.  Furthermore, even in the presence of such FSI's, one could
measure a ``true'' $T$-odd observable by measuring the TOPO in both
the $B$ and $\overline{B}$ modes and then taking the difference
in order to subtract out the FSI effects~\cite{OK}.
Ideally, one would measure both neutral and charged $B$ decays
in order to maximize the statistics.

\subsection{Form factors}
\label{sec:ff}

Let us begin by establishing some notation.  The relevant
hadronic matrix elements may be parameterized by the following form
factors,
\begin{mathletters}
\begin{eqnarray}
\langle D(p^{\prime})| \overline{c} \gamma_{\mu} b |B(p) \rangle
 & = & f_+ \, (p+p^{\prime})_{\mu}
      + f_- \, (p-p^{\prime})_{\mu} \\
\langle D(p^{\prime})| \overline{c} \gamma_{\mu} \gamma_5 b |B(p) \rangle
 & = & 0 \label{eq:axformfac}\\
\langle D^*(p^{\prime}, \epsilon)| \overline{c} \gamma^{\mu} b |B(p) \rangle
 & = &  i \frac{F_V}{m_B} \epsilon^{\mu\nu\alpha\beta}
         \epsilon^*_{\nu}  (p+ p^{\prime})_{\alpha} q_{\beta} \\
\langle D^*(p^{\prime}, \epsilon)| \overline{c} \gamma_{\mu}
   \gamma_5 b |B(p) \rangle
 & = & - F_{A0} \, m_B \epsilon^*_{\mu}
 -\frac{F_{A+}}{m_B} (p+p^{\prime})_{\mu} \epsilon^* \cdot q
 -\frac{F_{A-}}{m_B} q_{\mu} \epsilon^* \cdot q \, ,
\end{eqnarray}
\end{mathletters}
where $p$ and $p^{\prime}$ are the four-momenta of the $B$ and $D$
 ($D^*$) respectively, $\epsilon$ is the polarization
vector of the $D^*$, $q=p-p^{\prime}$, and the form factors
are functions of $q^2$.  We use the convention $\epsilon_{0123}=1$.
In the SM these form factors are relatively real to a good
approximation, but their functional dependences on $q^2$
are, a priori, unknown.
Note that the expression in Eq.~(\ref{eq:axformfac}) is equal to zero
since one cannot form an axial vector using only $p$ and $p^\prime$.

In order to derive the corresponding expressions for the scalar and
pseudoscalar hadronic matrix elements, we apply the
Dirac equation~\cite{cheng},
yielding
\begin{mathletters}
\begin{eqnarray}
\langle D(p^{\prime})| \overline{c} b |B(p) \rangle
& = & \frac{m_B^2}{m_b-m_c}[ f_+ \, (1-r_D) + f_- \, \frac{q^2}{m_B^2}] \\
\langle D(p^{\prime})| \overline{c} \gamma_5 b |B(p) \rangle
& =& 0 \\
\langle D^*(p^{\prime}, \epsilon)| \overline{c} b |B(p) \rangle
&=&0 \\
\langle D^*(p^{\prime}, \epsilon)| \overline{c} \gamma_5 b |B(p) \rangle
& =& \frac{m_B}{m_b+m_c} (\epsilon^* \cdot q)
    [ F_{A0} + F_{A+}\, (1-r_{D^*}) + F_{A-} \, \frac{q^2}{m_B^2}],
\end{eqnarray}
\end{mathletters}
where $m_b$ and $m_c$ are the masses of the $b$ and $c$ quarks,
$r_D=m_{D}^2/m_B^2$  and $r_{D^*}=m_{D^*}^2/m_B^2$.

There has been considerable progress in the past few years in
understanding the functional forms and interdependence of the
above form factors.  Isgur and Wise made the key observation in
1989~\cite{IW} that in the infinite mass limit
for the heavy quarks, all of the form factors are proportional
to each other and so may be expressed in terms of one universal
function, now called the Isgur-Wise function.  Corrections to this
picture due to the finite masses of the quarks, as well as
perturbative QCD effects,
can be incorporated in a systematic way in what has come to be
known as Heavy Quark Effective Theory (HQET)~\cite{neupr}.
In our numerical work, we will use the leading order results of HQET.
Our analytical results, however, will be written in terms of the
form factors themselves, with no assumptions about heavy quark symmetry.
In the heavy quark symmetry limit we have
\begin{mathletters}
\begin{eqnarray}
f_{\pm} & =& \pm\frac{1\pm\sqrt{r_{D}}}{2\sqrt[4]{r_{D}}} \xi(w) \\
F_V &=&  F_{A+} = - F_{A-} = \frac{1}{2\sqrt[4]{r_{D^*}}} \xi(w) \\
F_{A0} & =&- \sqrt[4]{r_{D^*}} (w+1) \xi(w) ,
\end{eqnarray}
\end{mathletters}
where $\xi$ denotes the Isgur-Wise function and
where $w=\frac{m_B^2+m_{D^{(*)}}^2-q^2}{2m_Bm_{D^{(*)}}}$.
The Isgur-Wise function is normalized to unity at zero recoil, $\xi(1)=1$.

It is convenient to parameterize the physics of semileptonic
$B$ decays in terms of effective four-Fermi interactions
as follows
\begin{eqnarray}
{\cal L}_{\rm eff} & = & - \frac{G_F}{\sqrt{2}} V_{cb}
        \overline{c} \gamma_{\alpha} (1 - \gamma_5) b
        \overline{\ell} \gamma^{\alpha} (1- \gamma_5) \nu
   +G_S \overline{c} b \overline{\ell} (1 - \gamma_5) \nu
   + G_P \overline{c} \gamma_5 b  \overline{\ell} (1 - \gamma_5) \nu
 \nonumber \\
   & &       + G_V \overline{c} \gamma_{\alpha}  b
         \overline{\ell} \gamma^{\alpha} (1- \gamma_5) \nu
            + G_A \overline{c} \gamma_{\alpha} \gamma_5 b
         \overline{\ell} \gamma^{\alpha} (1- \gamma_5) \nu
  + {\rm H.c.} , \label{eq:interaction}
\end{eqnarray}
where $G_F$ is the Fermi constant and $V_{cb}$ is the relevant
CKM matrix element.  The first term in the effective lagrangian
is due to the SM $W$-exchange diagram and the remaining terms
characterize contributions coming from new physics, with
$G_S$, $G_P$, $G_V$ and $G_A$ denoting the strengths of the new
effective scalar, pseudoscalar, vector and axial-vector
interactions, respectively.
The effects of effective tensor interactions are negligible
in most models and they will be omitted from the
present discussion for simplicity.
Note that since $T$ violation arises from the interference
between the SM amplitude,
which contains a left-handed neutrino, and the non-SM amplitude,
we do not need to consider four-Fermi operators involving a right-handed
neutrino.

 The new physics contributions to the decay amplitude
may be taken into account by the following
replacement of the form factors,
\begin{eqnarray}
f_+ & \rightarrow & f_+^{\prime}=f_+(1+\delta_+) \\
f_- & \rightarrow & f_-^{\prime}=f_-(1+\delta_-) \\
F_V &  \rightarrow & F^{\prime}_V =F_V(1+\delta_V) \\
F_{A0} &\rightarrow & F^{\prime}_{A0}=F_{A0}(1+\delta_{A0}) \\
F_{A+} &\rightarrow & F^{\prime}_{A+}=F_{A+}(1+\delta_{A+}) \\
F_{A-} &\rightarrow & F^{\prime}_{A-}=F_{A-}(1+\delta_{A-}) .
\end{eqnarray}
The $\delta$ parameters  are given by
\begin{eqnarray}
 \delta_+ & = & - \Delta_V  \\
 \delta_- & =& - \Delta_V  - \Delta_S \cdot
      \left[ \frac{f_+}{f_-} (1-r_D) + \frac{q^2}{m_B^2} \right] \\
 \delta_V & = &  -   \Delta_V \\
 \delta_{A0} & =& \Delta_A  \\
 \delta_{A+} & =& \Delta_A  \\
 \delta_{A-} & =&  \Delta_A  - \Delta_P \cdot
       \left[ \frac{F_{A0}}{F_{A-}}
         +\frac{F_{A+}}{F_{A-}} (1-r_{D^*})
         + \frac{q^2}{m_B^2} \right] ,
\end{eqnarray}
where
\begin{eqnarray}
\Delta_S & = & \frac{\sqrt{2}G_S}{G_FV_{cb}}  \frac{m_B^2}{(m_b-m_c)m_\ell}
	\label{eq:deltasdef} \\
\Delta_P & = & \frac{\sqrt{2}G_P}{G_FV_{cb}} \frac{m_B^2}{(m_b+m_c)m_\ell} \\
\Delta_V & = & \frac{\sqrt{2}G_V}{G_FV_{cb}} \\
\Delta_A & = & \frac{\sqrt{2}G_A}{G_FV_{cb}} \, .
\end{eqnarray}
  These $\delta$ ($\Delta$) parameters could in general
be complex and could then give rise to observable $C\!P$-violating effects.
Since it is typically true that the TOPO's which we will describe
are insensitive to new (SM-like) $V-A$ quark-current interactions, it is also
convenient to introduce one more parameter,
\begin{eqnarray}
\Delta_R & = & \frac{1}{2}(\Delta_V + \Delta_A) ,
\label{eq:deltaR}
\end{eqnarray}
which measures the strength of an effective right-handed quark-current
interaction.

\subsection{$\tau$ polarization in
$B \rightarrow D \tau \overline{\nu}$ decay}
\label{sec:BD}

Let us begin by deriving the expression for the $\tau$ lepton
transverse polarization in the semileptonic decay
\begin{eqnarray}
B(p) & \rightarrow & D(p^{\prime}) \tau(p_{\tau}) \overline{\nu}(p_{\nu}) .
\end{eqnarray}
The $\tau$ transverse
polarization in this decay is perfectly analogous to the muon
transverse polarization in $K_{\mu3}$ decay.
The amplitude arising from the general effective lagrangian of
Eq.~(\ref{eq:interaction}) can be written as
\begin{eqnarray}
{\cal M} & = & - \frac{G_F}{\sqrt{2}} V_{cb}
\overline{u}(p_{\tau})\gamma^{\mu} (1-\gamma_5) v(p_{\nu})
[ f_+^{\prime} \,  (p+p^{\prime})_{\mu} + f_-^{\prime} \,
(p-p^{\prime})_{\mu}] \, ,
\end{eqnarray}
which has the same form as the SM amplitude except for the replacement
$f_\pm$$\to$$f^\prime_\pm$.

The polarization observable may be written in
terms of two independent kinematical variables, which we
will take to be the energies of the $D$ meson and the $\tau$
lepton.  This choice is not unique, but is
convenient for our purposes.  Working in the $B$ rest frame,
we introduce dimensionless quantities $x$ and $y$ which
are proportional to these energies, but which are
normalized to half the $B$ mass,
$x$$=$$2p\cdot p^{\prime}/p^2$$=$$2E_{D}/m_B$ and
$y$$=$$2p\cdot p_{\tau}/p^2$$=$$2E_{\tau}/m_B$.
The differential partial width is then given by
\begin{eqnarray}
\frac{d^2\Gamma(B \rightarrow D \tau \overline{\nu})}{dxdy} & = &
 \frac{G_F^2 |V_{cb}|^2 m_B^5}{128 \pi^3}  \rho_D(x,y) \, ,
\end{eqnarray}
with
\begin{eqnarray}
  \rho_D(x,y) &  = &
   |f_+^{\prime}|^2 g_1(x,y) + 2{\rm Re}(f_+^{\prime}f_-^{\prime*})g_2(x,y)
  + |f_-^{\prime}|^2 g_3(x) \, .
\end{eqnarray}
The kinematical functions $g_i(x,y)$ are defined in Appendix~\ref{sec:appenda}.

  The transverse polarization of the $\tau$ lepton is then defined
as in Eq.~(\ref{eq:kmu3transpol}),
\begin{eqnarray}
P^{\bot(D)}_{\tau} & = & \frac{ d\Gamma({\vec n}) - d\Gamma(-{\vec n})}
                           { d\Gamma_{total}},
\end{eqnarray}
where $\vec{n}\equiv (\vec{p}_D \times \vec{p}_{\tau})/
      |\vec{p}_D \times \vec{p}_{\tau}|$ is a unit vector perpendicular to
the decay plane, and $d\Gamma(\pm\vec{n})$ is the differential partial
width with the $\tau$ spin vector along $\pm\vec{n}$.
$d\Gamma_{total}$ denotes the partial width after summing over the lepton
spins.  The above expression may be written in terms of $f_{\pm}^\prime$
as follows
\begin{eqnarray}
P^{\bot(D)}_{\tau}(x,y) & = & - \lambda_D(x,y)
{\rm Im}(2 f_+^{\prime} f_-^{\prime*}) ,
\end{eqnarray}
with
\begin{eqnarray}
\lambda_D(x,y) & = & \frac{\sqrt{r_{\tau}}}{\rho_D(x,y)}
 \sqrt{(x^2-4r_D)(y^2-4r_{\tau}) -4(1-x-y +\frac{1}{2}xy
        +r_D + r_{\tau})^2} \, ,
	\label{eq:lambdad}
\end{eqnarray}
where  $r_{\tau}=m_{\tau}^2/m_B^2$.

The expression for $P^{\bot(D)}_{\tau}$ can now be written
explicitly in terms of the effective four-Fermi interactions of
Eq.~(\ref{eq:interaction}) and then
simplified by keeping only the linear terms in the $\Delta$ parameters.
(The terms quadratic in $\Delta$ can easily be included if they are not
negligible in a specific model.)
This gives
\begin{mathletters}
\begin{eqnarray}
P^{\bot(D)}_{\tau}(x,y) & = & - \sigma_D(x,y) {\rm Im} \Delta_S  \\
\sigma_D(x,y) & = & h_D(x) \lambda_D(x,y) \\
h_D(x) & = & 2 f_+^2 (1-r_D) + 2 f_+ f_- (1-x+r_D) \, .
\end{eqnarray}
\end{mathletters}
To leading order in HQET the function $h_D(x)$ has a very simple
form, given by
\begin{eqnarray}
h_D(x) & \rightarrow & (1-r_D)
     \left( 1+ \frac{x}{2\sqrt{r_D}} \right) \xi^2 \, .
\label{eq:hD}
\end{eqnarray}

\begin{figure}[p]
\epsfsize=100pt \epsfbox[80 150 800 710]{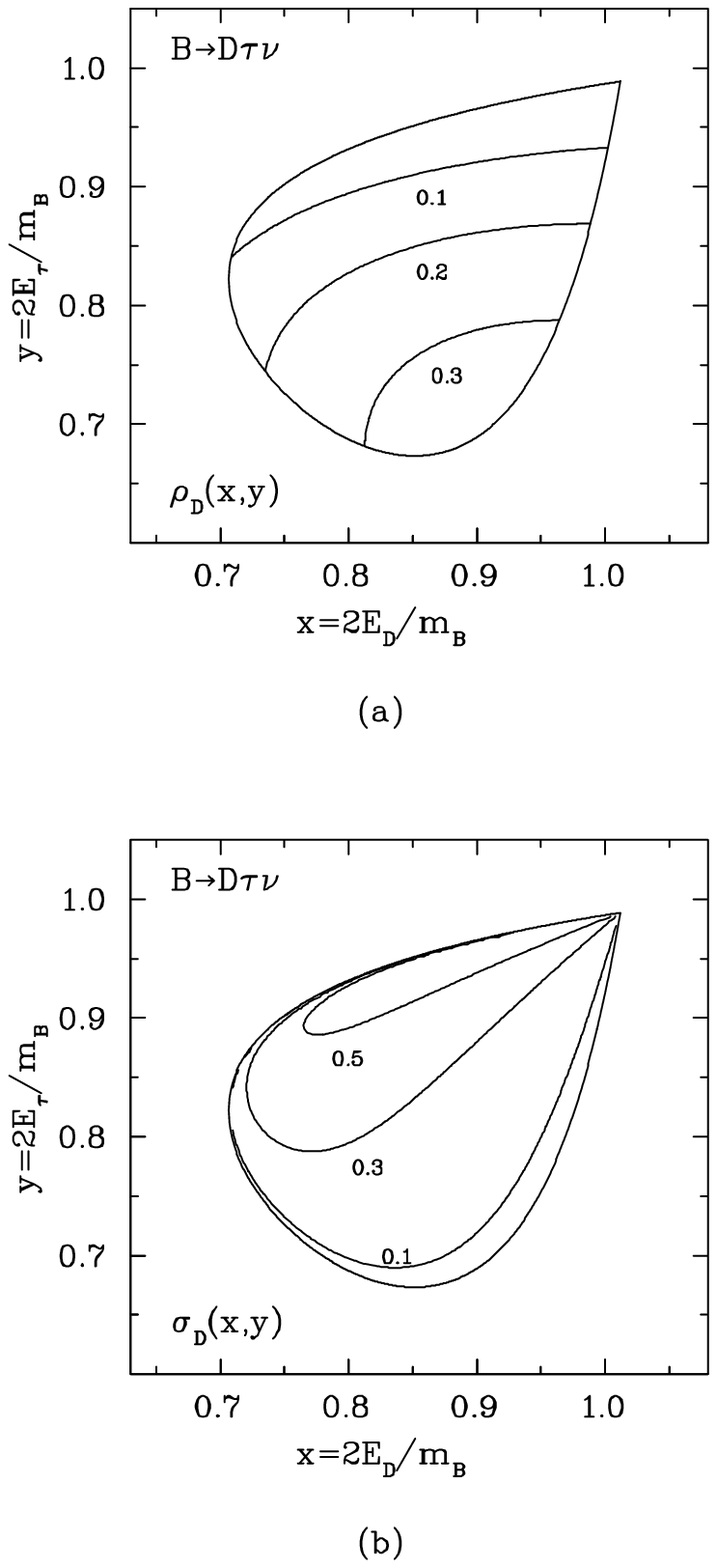}
\vspace{0pt}
\caption{Contour plots for the semileptonic decay
 $B \rightarrow D \tau \overline{\nu}$, using $\xi=1-0.75\times(w-1)$
 for the Isgur-Wise function: (a) the Dalitz density function
 $\rho_D(x,y)$; (b) the transverse $\tau$ polarization function
 $\sigma_D(x,y)$.}
\label{fig:BD}
\end{figure}

There are three features of these expressions which are of
interest.  First of all, note that the $\tau$ transverse polarization
in this decay is proportional to the effective
scalar four-Fermi interaction, as was claimed above.
This feature is well-known in the analogous $K_{\mu3}$ decay.
A second observation is that the polarization function
$\sigma_D(x,y)$ is explicitly proportional to the mass of the lepton
involved, and is therefore largest for the $\tau$ lepton.
The transverse polarization of the lepton will then be largest for
the $\tau$ mode in models for which $\Delta_S$ is independent of the
lepton mass\footnote{Note that the definition of $\Delta_S$ includes
a factor of $1/m_\ell$ which must be canceled in order for $\Delta_S$
to be independent of the lepton mass (see Eq.~(\ref{eq:deltasdef})).},
including multi-Higgs-doublet models and $R$-parity
conserving SUSY models with large intergenerational squark mixing.
If $\Delta_S$ depends on the lepton mass (as in, e.g.,
$R$-parity breaking SUSY models and leptoquark models), then
the lepton polarization need not be largest for the case $\ell$$=$$\tau$.
Our final observation is that, to leading order in HQET,
the Dalitz density $\rho_D(x,y)$ is proportional to
$\xi^2$, so that the polarization function $\sigma_D(x,y)$
is independent of $\xi(w)$.
The average polarization (defined below in Eq.~(\ref{eq:polav}))
does have a mild dependence on the form of the Isgur-Wise function.
This latter remark applies in general to polarization observables.
  The contour plots for $\rho_D(x,y)$ and $\sigma_D(x,y)$ are given
in Fig.~\ref{fig:BD}, taking $\xi(w)=1.0 -0.75\times(w-1)$, which
is representative of the current experimental data~\cite{dataxi}.

  The average polarization over a region of phase space $S$ can be defined as
follows
\begin{eqnarray}
\overline{P^{(D)}_{\tau}} & \equiv &
\frac{\int_S dxdy \rho_D(x,y) P^{\bot(D)}_{\tau}(x,y)}
     {\int_S dxdy \rho_D(x,y)} \, .
\label{eq:polav}
\end{eqnarray}
This average is a measure of the difference between the number
of $\tau$ leptons with their spins pointing above and below the decay plane
divided by the total number of $\tau$ leptons in the same region of
phase space $S$.
In terms of the four-Fermi interactions, we have
\begin{eqnarray}
\overline{P^{(D)}_{\tau}} & = &
- \overline{\sigma_D} {\rm Im} \Delta_S \, .
\end{eqnarray}
Since we are only keeping contributions to the polarization which
are first order in $\Delta_S$, $\overline{\sigma_D}$ is independent
of $\Delta_S$ and we may carry out the integration numerically.
Averaging over the whole phase space gives
\begin{eqnarray}
\overline{P^{(D)}_{\tau}} & = &
- 0.22 \times {\rm Im} \Delta_S \, .
\label{eq:avgtauD}
\end{eqnarray}

\subsection{$\tau$ polarization in
$B \rightarrow D^* \tau \overline{\nu}$ decay}
\label{sec:BD*}

The $\tau$ transverse polarization in the decay
\begin{eqnarray}
B(p) & \rightarrow & D^*(p^{\prime}) \tau(p_{\tau}) \overline{\nu}(p_{\nu})
\end{eqnarray}
is defined in complete analogy with that for the decay
to the $D$.  The general effective four-Fermi interactions  of
Eq.~(\ref{eq:interaction}) contribute to this decay
with an amplitude given by
\begin{eqnarray}
{\cal M} & = & - \frac{G_F}{\sqrt{2}} V_{cb}
\overline{u}(p_{\tau})\gamma_{\mu} (1-\gamma_5) v(p_{\nu})
\epsilon^*_{\rho} {\cal M}^{\rho\mu} \\
{\cal M}^{\rho\mu} & = & F_{A0}^{\prime} m_B g^{\rho\mu}
  +  \frac{F_{A+}^{\prime}}{m_B} (p+p^{\prime})^{\mu} q^{\rho}
  +  \frac{F_{A-}^{\prime}}{m_B} q^{\mu} q^{\rho}
  + i \frac{F_V^{\prime}}{m_B} \epsilon^{\mu\rho\alpha\beta}
           (p+ p^{\prime})_{\alpha} q_{\beta} \, .
\label{eq:mrhomu}
\end{eqnarray}

Working again in the $B$ rest frame, we define
$x$$=$$2p\cdot p^{\prime}/p^2$$=$$2E_{D^*}/m_B$ and
$y$$=$$2p\cdot p_{\tau}/p^2$$=$$2E_{\tau}/m_B$.
Summing over the spins of the final states, we
find the following expression for the
differential partial width:
\begin{eqnarray}
\frac{d^2\Gamma(B \rightarrow D^* \tau \overline{\nu})}{dxdy} & = &
 \frac{G_F^2 |V_{cb}|^2 m_B^5}{128 \pi^3}  \rho_{D^*}(x,y) ,
\end{eqnarray}
with
\begin{eqnarray}
  \rho_{D^*}(x,y) &  = &
    |F_{A0}^{\prime}|^2 f_1(x,y) + |F_{A+}^{\prime}|^2 f_2(x,y) +
     |F_{A-}^{\prime}|^2 f_3(x)
      +   |F_V^{\prime}|^2  f_4(x,y)
 \nonumber \\ &&
      + 2 \rm{Re}(F_{A0}^{\prime} F_{A+}^{\prime\,*}) f_{5}(x,y)
      + 2 \rm{Re}(F_{A0}^{\prime} F_{A-}^{\prime\,*}) f_{6}(x,y)
 \nonumber \\ &&
      + 2 \rm{Re}(F_{A+}^{\prime} F_{A-}^{\prime\,*}) f_{7}(x,y)
      + 2 \rm{Re}(F_{A0}^{\prime} F_V^{\prime\,*}) f_{8}(x,y) \, .
\label{eq:rhoD*}
\end{eqnarray}
The subscripts of the eight functions denote the corresponding
contributions from the different form factors. These functions are
collected in Appendix~\ref{sec:appenda}.

   After a kinematic analysis, it is found that only interference
terms between the axial form factors contribute to
$P^{\bot(D^*)}_{\tau}$, so that
\begin{eqnarray}
P^{\bot(D^*)}_{\tau} & = &
   -  \lambda_{D^*}(x,y)
 \left[
   {\rm Im}(F_{A0}^{\prime} F_{A+}^{\prime\,*}) \left( \frac{x}{2r_{D^*}}
		+1 \right)
 + {\rm Im}(F_{A0}^{\prime} F_{A-}^{\prime\,*}) \left( \frac{x}{2r_{D^*}}
		-1 \right)
 \nonumber  \right. \\
 && \left.\!\! + {\rm Im}(F_{A+}^{\prime} F_{A-}^{\prime\,*})
    \left( \frac{x^2}{2r_{D^*}} -2 \right) \right] \, ,
\end{eqnarray}
with
\begin{eqnarray}
\lambda_{D^*}(x,y) & = &  \frac{\sqrt{r_{\tau}}}{\rho_{D^*}(x,y)}
 \sqrt{(x^2-4r_{D^*})(y^2-4r_{\tau}) -4(1-x-y +\frac{1}{2}xy
        +r_{D^*} + r_{\tau})^2} \, .
\end{eqnarray}

The $\tau$ transverse polarization may now be written
in terms of the effective four-Fermi interactions of
Eq.~(\ref{eq:interaction}).  Keeping the leading, linear terms
in the $\Delta$ parameters, we find
\begin{mathletters}
\begin{eqnarray}
P^{\bot(D^*)}_{\tau} & = & - \sigma_{D^*}(x,y) {\rm Im} \Delta_P
\label{eq:43a} \\
\sigma_{D^*}(x,y) & = & h_{D^*}(x) \lambda_{D^*}(x,y)  \\
h_{D^*}(x) & = & [ F_{A0} + F_{A+} (1-r_{D^*}) + F_{A-} (1-x+r_{D^*})]
\nonumber \\
            && \times   \left[ F_{A0} \left(\frac{x}{2r_{D^*}} -1\right)
               + F_{A+} \left(\frac{x^2}{2r_{D^*}} -2\right) \right] \, .
\label{eq:hdstar}
\end{eqnarray}
\end{mathletters}
The expression for $h_{D^*}(x)$ again has a very simple
form in the heavy quark symmetry limit,
\begin{eqnarray}
h_{D^*}(x) & \rightarrow & (1-r_{D^*})
	\left( 1+ \frac{x}{2 \sqrt{r_{D^*}}} \right)\xi^2 \, .
\end{eqnarray}
Comparison with Eq.~(\ref{eq:hD}) shows that this expression
may be obtained from the analogous expression for $h_D(x)$
by taking $r_D$$\rightarrow$$r_{D^*}$.

The $\tau$ lepton polarization in the
$B \rightarrow D^* \tau \overline{\nu}$ decay
is sensitive only to effective pseudoscalar
four-Fermi interactions~\cite{wkn}.  This observable is
thus complementary to its analogue in the decay
$B \rightarrow D \tau \overline{\nu}$, which is
sensitive to effective scalar interactions.
We note in passing that Garisto~\cite{garisto} has found
the transverse tau polarization in
$B \rightarrow D^* \tau \overline{\nu}$ to have an additional
dependence on effective right-handed quark-current interactions.
There is no discrepancy with our results, however,
since the effect which Garisto discusses only
arises when one fixes the polarization state of the $D^*$, instead of
summing over polarizations as we have done.  The right-handed current
effect cancels in the sum.  In the next subsection we will discuss
an observable which is sensitive to such right-handed interactions
but which requires a measurement of {\em only} the $D^*$ polarization
(not that of both the $D^*$ and the $\tau$.)

\begin{figure}[p]
\epsfsize=100pt \epsfbox[80 150 800 710]{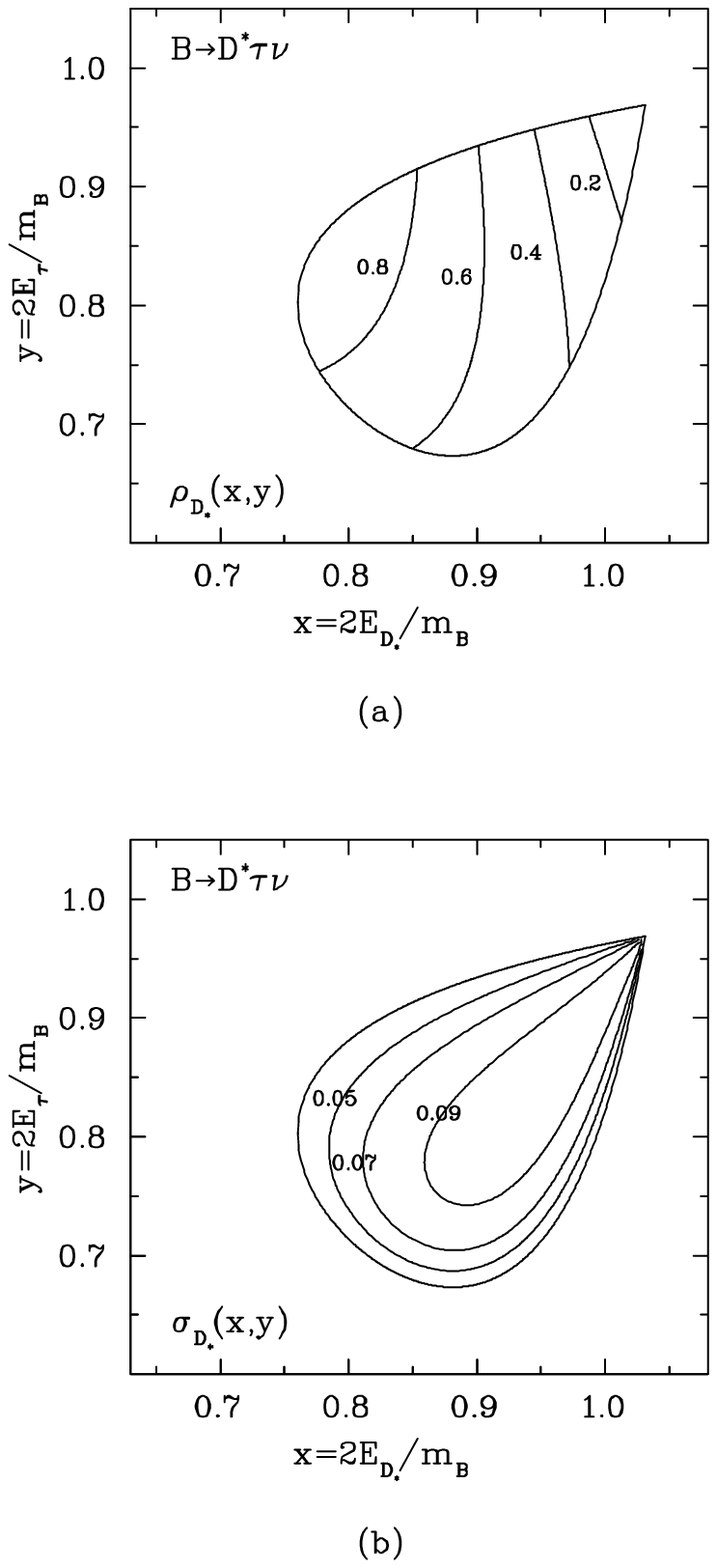}
\vspace{0pt}
\caption{Contour plots for the semileptonic decay
 $B \rightarrow D^* \tau \overline{\nu}$, using $\xi=1-0.75\times(w-1)$
 for the Isgur-Wise function: (a) the Dalitz density function
 $\rho_{D^*}(x,y)$; (b) the transverse $\tau$ polarization function
 $\sigma_{D^*}(x,y)$.}
\label{fig:BD*}
\end{figure}

 As noted in section~\ref{sec:BD},
the Dalitz density  $\rho_{D^*}(x,y)$ is quadratically dependent on
$\xi(w)$, whereas the polarization function
$\sigma_{D^*}(x,y)$ is to a good approximation independent
of $\xi(w)$. The average polarization varies slightly with $\xi(w)$.
 The contour plots for $\rho_{D^*}(x,y)$ and $\sigma_{D^*}(x,y)$
are given in Fig.~\ref{fig:BD*},  taking  again
$\xi(w)=1.0 - 0.75 \times (w-1)$ \cite{dataxi}.

  The average transverse polarization of the $\tau$ lepton can be defined
as in Eq.~(\ref{eq:polav}).  Averaging over the whole phase space gives
\begin{eqnarray}
\overline{P^{(D^*)}_{\tau}} & = &
- \overline{\sigma_{D^*}} {\rm Im} \Delta_P
 = - 0.068 \times {\rm Im} \Delta_P \, .
\label{eq:avgtauDst}
\end{eqnarray}
Note that $\overline{\sigma_{D^*}}$ is about a factor of three
smaller than $\overline{\sigma_{D}}$.
 This is because effectively only one of the three polarization states
of the $D^*$, the longitudinal polarization, contributes to the
transverse $\tau$ polarization \cite{wkn}.

\subsection{$D^*$ polarization in
$B \rightarrow D^* \ell \overline{\nu}$ decay}
\label{sec:D*pol}

In the previous two subsections we have looked at TOPO's constructed
using the spin of the tau in the decays
$B$$\rightarrow$$D\ell\overline{\nu}$ and
$B$$\rightarrow$$D^*\ell\overline{\nu}$.  Since the $D^*$
is a vector meson, however, the latter channel offers additional
TOPO's which may be constructed by using the projection
of the $D^*$ polarization transverse to the decay plane.
As we have already noted, these new observables will
be sensitive to effective right-handed current interactions, making
them complementary to the lepton transverse
polarization observables discussed above.

Let us denote the three-momenta of the $D^*$ and $\ell$
in the $B$ rest frame by $\vec{p}_{D^*}$ and $\vec{p}_{\ell}$,
respectively.  We may then define three orthogonal
vectors ${\vec n}_1$, ${\vec n}_2$, and ${\vec n}_3$ by
\begin{mathletters}
\begin{eqnarray}
{\vec n}_1 & \equiv & \frac{(\vec{p}_{D^*} \times \vec{p}_{\ell})
    \times \vec{p}_{D^*}}
     {|(\vec{p}_{D^*} \times \vec{p}_{\ell})
    \times \vec{p}_{D^*}|} \label{eq:n1}\\
{\vec n}_2 & \equiv & \frac{\vec{p}_{D^*} \times \vec{p}_{\ell}}
                     {|\vec{p}_{D^*} \times \vec{p}_{\ell}|}\label{eq:n2}
  \\
{\vec n}_3 & \equiv & \frac{\vec{p}_{D^*}}{|\vec{p}_{D^*}|}
      \frac{m_{D^*}}{E_{D^*}} \, .
\label{eq:n3}
\end{eqnarray}
\end{mathletters}
The unusual normalization of $\vec{n}_3$ is due to the boost from the
$D^*$ rest frame to the $B$ rest frame.
The constraint $\epsilon^2=-1$ can now be written in a symmetric form,
\begin{eqnarray}
  ({\vec \epsilon}\cdot {\vec n}_1)^2 +
  ({\vec \epsilon}\cdot {\vec n}_2)^2 +
   ({\vec \epsilon}\cdot {\vec n}_3)^2 & = & 1 .
\end{eqnarray}
 Note that ${\vec n}_1$ and ${\vec n}_3$ lie in the decay plane,
whereas ${\vec n}_2$ is perpendicular to the decay plane.

   The polarization vector of the $D^*$ can be taken to be real.
It is then clear from Eqs. (\ref{eq:n1}) - (\ref{eq:n3}) that
the $D^*$ polarization projection transverse to the decay plane,
${\vec \epsilon}\cdot {\vec n}_2$, is $T$-odd, and that
the $D^*$ polarization projections inside the decay plane,
${\vec \epsilon}\cdot {\vec n}_1$ and ${\vec \epsilon}\cdot {\vec n}_3$,
are $T$-even.  Since the polarization vector always comes
up quadratically in the differential width, the pieces which are
odd under time reversal must be proportional to
$({\vec \epsilon}\cdot {\vec n}_2)({\vec \epsilon}\cdot {\vec n}_1)$
or to $({\vec \epsilon}\cdot {\vec n}_2)({\vec \epsilon}\cdot {\vec n}_3)$.
For the moment we will define observables explicitly in terms
of the $D^*$ polarization vector.  At the end of this subsection
we will comment on how one could measure these quantities by
measuring the angular distributions of the decay products of the $D^*$.

Let us then formally define a measure of the $T$-odd
correlation involving the $D^*$ polarization as follows
\begin{eqnarray}
P^{(\ell)}_{D^*} & \equiv &
 \frac{d\Gamma - d\Gamma^{\prime}}{d\Gamma_{total}} =
  \frac{2d\Gamma_{T-odd}}{d\Gamma_{total}} ,
\label{eq:polD*}
\end{eqnarray}
where $d\Gamma^{\prime}$ is obtained by performing a $T$-transformation
on $d\Gamma$, $d\Gamma_{T-odd}$ is the $T$-odd piece in the partial width,
and $d\Gamma_{total}$ is the partial width after summing over
polarizations in the final state.  Note that there is also an implicit
sum over the spin of the final state charged lepton in Eq.~(\ref{eq:polD*}),
so that this observable depends {\em only} on the $D^*$ polarization.
We may then express this observable in terms of the two independent
kinematical variables $x$ and $y$, yielding
\begin{eqnarray}
P^{(\ell)}_{D^*}(x,y) & = &
  - ({\vec \epsilon}\cdot {\vec n}_1)({\vec \epsilon}\cdot {\vec n}_2)
   \lambda_1(x,y)  {\rm Im}(F_{A0}^{\prime} F_V^{\prime\,*})
  \nonumber  \\ &&
 + ({\vec \epsilon}\cdot {\vec n}_3)({\vec \epsilon}\cdot {\vec n}_2)
 \lambda_2(x,y) [
    {\rm Im}(F^{\prime}_{A0} F_{A+}^{\prime\,*})
    +{\rm Im}(F_{A+}^{\prime} F_V^{\prime\,*}) (x+2y-2-r_\ell)
  \nonumber  \\ && ~~~~~~~~~~~~~~~~~~~~~~~~~~~~~
   +{\rm Im}(F_{A-}^{\prime} F_V^{\prime\,*}) r_\ell
    +{\rm Im}(F_{A0}^{\prime} F_V^{\prime\,*}) d_\ell(x,y)] ,
\label{eq:poldstarl}
\end{eqnarray}
with
\begin{mathletters}
\begin{eqnarray}
 \lambda_1(x,y) & = &  \frac{
    4 [(x^2-4r_{D^*})(y^2-4r_{\ell}) -4(1-x-y +\frac{1}{2}xy
        +r_{D^*} + r_{\ell})^2]}{\rho_{D^*}(x,y)\sqrt{x^2-4r_{D^*}}} \\
 \lambda_2(x,y) & = &
\frac{4\sqrt{\frac{x^2}{4r_{D^*}}-1}}{\rho_{D^*}(x,y)}
 \sqrt{(x^2-4r_{D^*})(y^2-4r_{\ell}) -4(1-x-y +\frac{1}{2}xy
        +r_{D^*} + r_{\ell})^2} \\
d_\ell(x,y) & = & (y-1)
    - \frac{2x(1+r_{D^*}+r_\ell-x-y+\frac{1}{2}xy)}{x^2-4r_{D^*}} \, ,
\end{eqnarray}
\end{mathletters}
where
$r_\ell=m_\ell^2/m_B^2$, with $\ell$$=$$e$, $\mu$, $\tau$.

These expressions may be simplified by writing them in terms
of the effective four-Fermi interactions of
Eq.~(\ref{eq:interaction}) and neglecting terms quadratic in
the $\Delta$ parameters.  This gives
\begin{eqnarray}
P^{(\ell)}_{D^*}(x,y) & = &
   ({\vec \epsilon}\cdot {\vec n}_1)({\vec \epsilon}\cdot {\vec n}_2)
   \sigma^{\ell}_1(x,y) {\rm Im} \Delta_R
\nonumber \\ && +
 ({\vec \epsilon}\cdot {\vec n}_3)({\vec \epsilon}\cdot {\vec n}_2)
 [\sigma^{\ell}_2(x,y) {\rm Im} \Delta_R
+ \sigma^{\ell}_3(x,y) {\rm Im} \Delta_P] \, ,
\label{eq:pdstartau}
\end{eqnarray}
where
\begin{mathletters}
\begin{eqnarray}
\sigma^{\ell}_1(x,y) & = & -2 \lambda_1(x,y) F_{A0} F_V  \\
\sigma^{\ell}_2(x,y) & = & 2 \lambda_2(x,y) F_V [
      F_{A+} (x+2y-2-r_\ell) +F_{A-} r_\ell
        + F_{A0} d_\ell(x,y) ]
  \\
\sigma^{\ell}_3(x,y) & = & - \lambda_2(x,y) r_\ell
      F_{V} [ F_{A0} + F_{A+} (1-r_{D^*}) + F_{A-} (1+r_{D^*}-x)] .
\end{eqnarray}
\end{mathletters}
The $T$-odd $D^*$ polarization observable can thus receive
contributions from both right-handed current and effective pseudoscalar
interactions.  The pseudoscalar contribution is
suppressed by $r_\ell$, however, so that
the decay modes $B \rightarrow D^* e \overline{\nu}$
and $B \rightarrow D^* \mu \overline{\nu}$ may be used to isolate and measure
the right-handed current effect.
 As we have noted above in sections~\ref{sec:BD} and \ref{sec:BD*},
the transverse polarization of the $\tau$ lepton is sensitive to
an effective scalar four-Fermi interaction in the decay
$B \rightarrow D \tau \overline{\nu}$, and to
a pseudoscalar interaction in the decay
$B \rightarrow D^* \tau \overline{\nu}$.
Combining all three polarization measurements,
it is thus possible to probe separately the three different sources of
non-standard model $T$-violation which we have included in the effective
lagrangian of Eq.~(\ref{eq:interaction})\footnote{We note again that
$T$-odd observables are typically
insensitive to new {\em left}-handed interactions
since the interference of such diagrams with the SM diagram does not
lead to observable phases.}.

\begin{figure}[p]
\epsfsize=100pt \epsfbox[80 150 800 710]{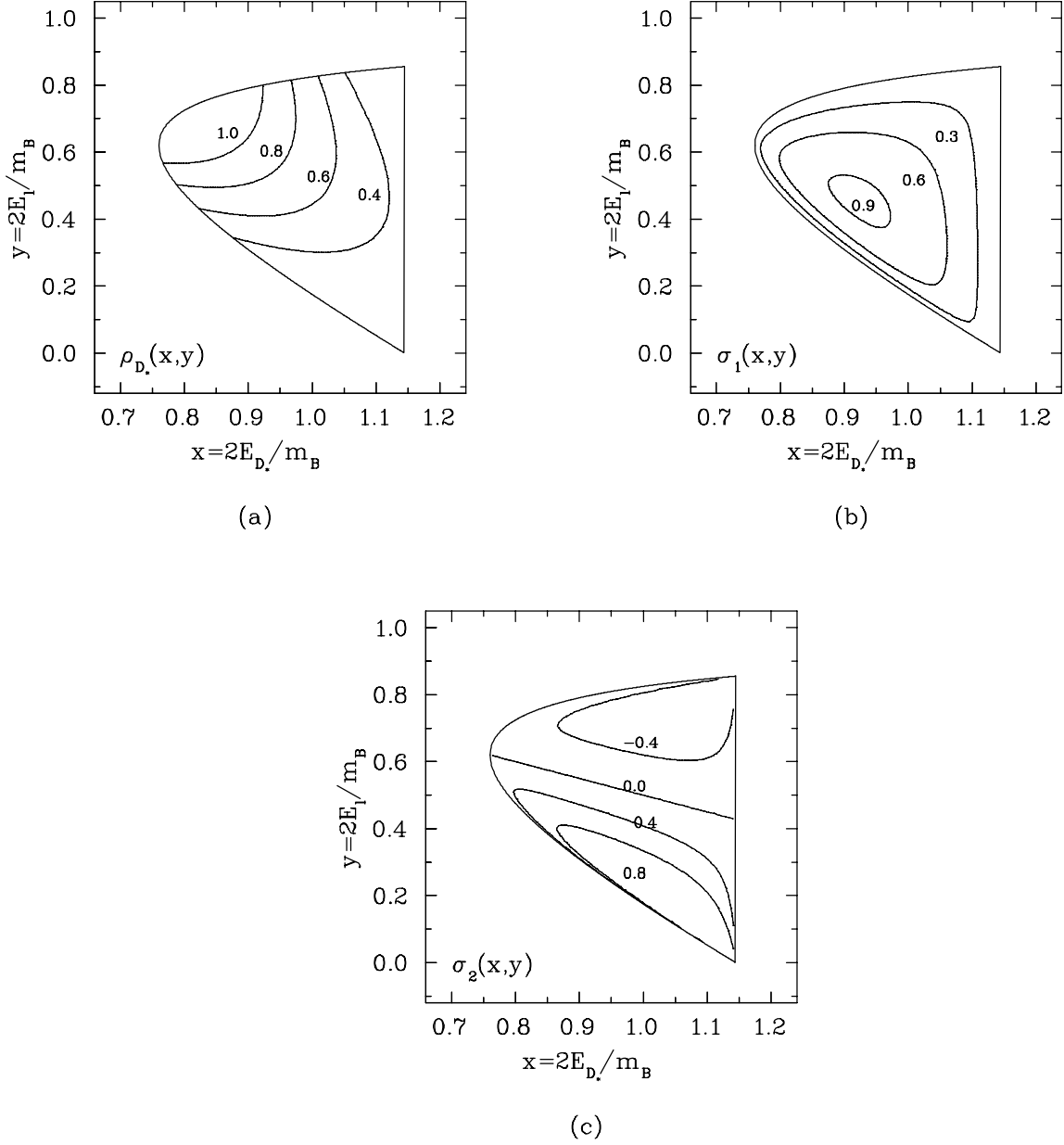}
\vspace{0pt}
\caption{Contour plots for the semileptonic decay
 $B \rightarrow D^* \ell \overline{\nu}$ ($\ell$$=$$e$,$\mu$), using
$\xi=1-0.75\times(w-1)$
 for the Isgur-Wise function: (a) the Dalitz density function
 $\rho^{\ell}_{D^*}(x,y)$; (b) the $D^*$ polarization function
 $\sigma_1(x,y)$; (c) the $D^*$ polarization function $\sigma_2(x,y)$.
  The masses of the leptons are neglected in these plots.}
\label{fig:D*pol}
\end{figure}

For the remainder of this section we will concentrate on
the $\ell$$=$$e$ and $\mu$ modes, studying
their sensitivity to an effective right-handed current interaction.
Aside from the fact that these two channels naturally isolate
the effective right-handed interactions, they are also favored by
virtue of their larger branching fractions compared to $\ell$$=$$\tau$.
Given the small masses of the electron and muon compared to the other
energy scales in the problem, we may safely set
$r_{\ell}$$=$$m_{\ell}^2/m_B^2$$=$$0$.  We will subsequently also
drop the superscript $\ell$.  The expression for the $T$-odd
$D^*$ polarization observable then becomes
\begin{eqnarray}
P_{D^*}(x,y) & = &
  [({\vec \epsilon}\cdot {\vec n}_1) \sigma_1(x,y)
  + ({\vec \epsilon}\cdot {\vec n}_3)\sigma_2(x,y)]
({\vec \epsilon}\cdot {\vec n}_2) {\rm Im} \Delta_R \, ,
\label{eq:D*pol}
\end{eqnarray}
where, to leading order in HQET, the two polarization functions
are given by
\begin{mathletters}
\begin{eqnarray}
\sigma_1(x,y) &\rightarrow  &  \lambda_1(x,y)(x+2\sqrt{r_D})
\frac{\xi^2}{2 \sqrt{r_D}} \\
\sigma_2(x,y) & \rightarrow &
\lambda_2(x,y) (2-x-2y)\frac{1-\sqrt{r_D}}{x-2\sqrt{r_D}} \xi^2
\end{eqnarray}
\end{mathletters}
Note that both $\lambda_1(x,y)$ and $\lambda_2(x,y)$ are proportional
to $1/\rho_{D^*}$, and therefore to $1/\xi^2$.
The polarization functions $\sigma_i(x,y)$ ($i=1,2$) are
then independent of the Isgur-Wise function $\xi(w)$ as noted above.
   The contour plots for the Dalitz density function $\rho_{D^*}(x,y)$
 and the polarization functions
$\sigma_1(x,y)$ and $\sigma_2(x,y)$
are shown in Fig.~\ref{fig:D*pol}, assuming
$\xi(w)=1.0 -0.75 \times (w-1)$ \cite{dataxi}.

We have previously analyzed the two different polarization
structures present in the expression for $P_{D^*}(x,y)$~\cite{wkn}.
As was noted there, the term proportional to $\sigma_1$ involves
only transverse polarization components, while that proportional
to $\sigma_2$ requires a non-zero longitudinal projection of
the $D^*$ polarization in order to be non-vanishing\footnote{
This latter term would be absent for on-shell massless vector
bosons such as the photon.}.  In addition to multiplying distinct
polarization structures, however, the two functions $\sigma_1$
and $\sigma_2$ themselves have quite different
symmetry properties in the two-dimensional
phase space spanned by $x$ and $y$.  In principle, then, there
are at least two distinct ways in which to differentiate between
the two contributions to $P_{D^*}(x,y)$.  The first is to devise
a method which can pick out one or the other polarization
structure, and the second is to make use of the symmetries of $\sigma_1$
and $\sigma_2$ in order to differentiate between them.  The latter of
these two has been discussed in some detail in Ref.~\cite{wkn}, so let us
first recapitulate those results and then discuss how one can get at the
polarization structures themselves.

It is straightforward to demonstrate that
$\rho_{D^*}(x,y)\sigma_2(x,y)$ is antisymmetric under the
exchange of lepton and anti-neutrino energies, and that the allowed
phase space region is symmetric under the same
exchange.  Thus, integrating over all of phase space -- or over
any region which is symmetric under the exchange -- eliminates
the $\sigma_2$ term and leaves only the piece due to the
$\sigma_1$ term.  In order to pick out $\sigma_2$, we note
that $\rho_{D^*}\sigma_1$ is symmetric under $y \to 2-x-y$ and $x \to x$,
so that an asymmetric average over phase space may be used to
eliminate the $\sigma_1$ term.  In both cases, these properties
are independent of the functional forms of the form factors.
Performing these averages over all phase space then yields
for the non-vanishing piece in the two cases
\begin{eqnarray}
  \overline{P^{(1)}_{D^*}} &\simeq & 0.51 \times
({\vec \epsilon}\cdot {\vec n}_1)({\vec \epsilon}\cdot {\vec n}_2)
{\rm Im} \Delta_R \, , \label{eq:TOPO1} \\
  \overline{P^{(2)}_{D^*}}
& \equiv & \frac{ \int dx \left( \int_{y_{min}}^{y_{mid}} dy
                               -\int_{y_{mid}}^{y_{max}} dy \right)
     \rho_{D^*}(x,y) P_{D^*}(x,y) }
    {\int dx dy \rho_{D^*}(x,y)}  \nonumber \\
 &\simeq & 0.40 \times
(\vec{\epsilon}\cdot {\vec{n}}_2)(\vec{\epsilon}\cdot \vec{n}_3)
{\rm Im} \Delta_R ,
\label{eq:TOPO2}
\end{eqnarray}
where $y_{mid}$$\equiv$$(y_{min} + y_{max})/2$.
The asymmetric-average approach used to pick out the $\sigma_2$ term
in Eq.~(\ref{eq:TOPO2}) also works when the lepton is not massless.
In fact, this method also eliminates the extra pseudoscalar term
which is present in Eq.~(\ref{eq:pdstartau}), so that even for $\ell$$=$$\tau$
it is possible to isolate the right-handed current contribution.
Numerically, however, the average $D^*$ polarization found using
this prescription is about a factor of three smaller for the tau
compared to the electron and muon channels.

An alternative method for differentiating between the two polarization
structures is to examine the angular distributions of the decay
products of the vector meson.  It is straightforward to demonstrate
that the resulting asymmetries (integrated appropriately over the momenta
of the final state particles) have the same structure as the terms
which define $P_{D^*}^{(\ell)}(x,y)$ in Eq.~(\ref{eq:pdstartau}),
up to the replacement of the factors
$(\vec{\epsilon}\cdot\vec{n}_i)(\vec{\epsilon}\cdot\vec{n}_j)$ by
a numerical factor of $1/\pi$.  In Appendix~\ref{sec:appendb},
we demonstrate this explicitly for the decay mode
$D^*$$\to$$D\pi$\footnote{
In the case of the neutral $D^*$, and depending
on the experimental set-up, it might be easier to use the
$D^*$$\to$$D\gamma$ mode, since the two photons
 from the $\pi^0$ decay could be quite soft.  The charged pions produced
in the case of charged $D^*$ decays, however, should be easier
to detect~\cite{coward}.}.

   Before we turn to the section on model estimates, it is worth pointing
out that the numerical
coefficients in Eqs.~(\ref{eq:avgtauD}), (\ref{eq:avgtauDst}),
(\ref{eq:TOPO1}) and (\ref{eq:TOPO2}), evaluated at leading order
in the heavy quark expansion,
will be modified when the effects due to finite quark masses and QCD
corrections are included. The uncertainty in the Isgur-Wise function
can also affect these coefficients but to a much lesser extent, as mentioned
earlier.  In order to get a feel for the size of these corrections,
we have reevaluated the coefficients using the QCD sum rule
estimates for $\xi(w)$ and for the form factors given in Table~5.1
of Ref.~\cite{neupr}.  The estimates in Ref.~\cite{neupr}
correspond to next-to-leading order in the $1/m_Q$ expansion of HQET.
Our findings are that the correction to the $\tau$ polarization
in $B \rightarrow D \tau \overline{\nu}$ is less than one percent
of the value quoted in Eq.~(\ref{eq:avgtauD}), while 
the analogous correction for 
$B \rightarrow D^* \tau \overline{\nu}$ leads to an increase 
of about $15\%$ in the magnitude of the polarization.
For the $D^*$ polarization in $B \rightarrow D^* \ell \overline{\nu}$
($\ell=e,\mu$),
$\overline{P^{(1)}_{D^*}}$ and $\overline{P^{(2)}_{D^*}}$
were found to increase respectively by about $20\%$ and $25\%$
relative to the values quoted in Eqs.~(\ref{eq:TOPO1}) and (\ref{eq:TOPO2}).
Considering the uncertainties in our current knowledge of the form factors,
we will simply use the leading order results obtained in this 
section when making our model estimates.
It should be understood, however, that more precise knowledge of the
form factors could change our estimates (generally increasing them) 
by up to about $25\%$.

   The main results of our general analysis are listed in
Table~\ref{tab:general}.

\section{Model Estimates}
\label{sec:models}

In this section we examine the prospects for the various $T$-odd
observables, both in supersymmetric models and in some non-supersymmetric
models.  We start by looking at SUSY models that conserve $R$-parity.
In this case, there are no $C\!P$-violating contributions
to our observables at tree-level.
As we have noted elsewhere, however \cite{wkn}, there can be
rather large effects (even though they occur at one loop)
in SUSY models with intergenerational squark mixings.  In this case
both the $\tau$ polarization and $D^*$ polarization observables can receive
sizeable contributions.  We then examine models in which
$R$-parity is explicitly violated.  In such models, $T$-violating
scalar and pseudoscalar interactions can arise
at tree level, leading to non-zero values for the
transverse $\tau$ polarization in the decays
$B \rightarrow D^{(*)} \tau \overline{\nu}$.
The sizes of these observables are subject to stringent experimental
constraints.  We next consider some non-SUSY extensions of the SM.
We first examine the multi-Higgs-doublet and leptoquark models, which
can both induce effective scalar and pseudoscalar four-Fermi interactions
at tree-level. Then we look at left-right symmetric models, where
we focus on the effects due to the extra gauge bosons only, and give
an estimate of the size of the $T$-odd $D^*$ polarization observable.

The results obtained in these models are summarized in Table~\ref{tab:models}.

\subsection{SUSY with Intergenerational Squark Mixing}
\label{sec:QSM}

  The notion of squark family mixings comes
 from the observation that the mass matrices of the
quarks and squarks are generally expected to be
 diagonalized by different unitary
transformations in generation space \cite{fcnc,fcncm1,fcncm2}.
The relative flavor rotations
between the $\tilde{u}_L$, $\tilde{u}_R$,
$\tilde{d}_L$, and $\tilde{d}_R$ squarks and their corresponding
quark partners are denoted by the three by three unitary matrices
$V^{U_L}$, $V^{U_R}$, $V^{D_L}$, and $V^{D_R}$, respectively.
The significance of these mixings for $T$-violating semileptonic
meson decays was noted in a previous work \cite{wn1} and discussed
in some detail for the transverse muon polarization in $K^+_{\mu3}$ \cite{wn1}
and $K^+_{\mu2\gamma}$ \cite{wn2} decays.
In this subsection, we focus on the various TOPO's in different
exclusive semileptonic $B$ decay channels \cite{wkn}.

 To estimate the maximal $T$-violation effects in semileptonic
$B$ decays, we consider the one-loop diagrams \cite{wn1,wn2} with a
gluino ($\tilde{g}$) and top and bottom squarks ($\tilde{t}$, $\tilde{b}$)
in the loop, and with $W$ or charged Higgs exchange.
The relevant mixing matrix elements involved
are $V^{U}_{32}$ and $V^{D}_{33}$.
When the mixing is large, we can have
doubly-enhanced $T$-violation effects -- due to mixing and to the
large top quark mass.
Note that flavor changing neutral current processes
 only constrain the combinations
$V^U {V^U}^*$ and $V^D {V^D}^*$.
For example, $D$-$\bar{D}$ mixing can put nontrivial constraints
on the product $V^{U}_{32} {V^{U}_{31}}^*$.
We will assume maximal mixing between the
($\tilde{t}_R$, $\tilde{c}_R$) squarks and thus take
$|V^{U_R}_{32}|=\frac{1}{\sqrt{2}}$ to estimate
the maximal polarization effects.

\subsubsection{$H^+$ exchange and $\tau$ lepton polarization}
\label{sec:H}

Charged Higgs exchange can give rise to effective scalar and pseudoscalar
but not vector and axial-vector interactions, as can be seen from
Lorentz invariance of the amplitude.
It could thus contribute to the transverse
polarization of the $\tau$ lepton in both
$B \rightarrow D \tau \overline{\nu}$ and
$B \rightarrow D^* \tau \overline{\nu}$ decays,
but it does not contribute  to the $D^*$ polarization in the $e,\mu$ modes.
 Furthermore, in the large $\tan \beta$ limit, the
 induced effective scalar and pseudoscalar
 interactions from $W$-boson exchange
are suppressed by $1/\tan \beta$ relative to the
 charged Higgs exchange.
  To estimate the maximal size of $P^{\bot}_{\tau}$, we need
 only consider charged Higgs contributions.

The $m_t$-enhanced effective scalar-pseudoscalar
four-Fermi interaction can be estimated
 from the diagram that contains a $\tilde{g}$-$\tilde{t}$-$\tilde{b}$
loop and the $H^- \tilde{t}_R {\tilde{b}_L}^*$ vertex.
It is given by \cite{wn2,wkn}
\begin{eqnarray}
{\cal L}_H & = & \frac{4G_F}{\sqrt{2}} C_H
(\overline{c}_R b_L)(\overline{\tau}_R \nu_L)
 + \rm{H.c.} ,
\label{eq:LH}
\end{eqnarray}
with
\begin{eqnarray}
C_H & = & - \frac{\alpha_s}{3\pi} I_H \tan \beta
 \frac{m_t m_{\tau}}{m_H^2}
\frac{\mu + A_t\cot\beta}{m_{\tilde{g}}}
V^{H}_{33} V^{D_L}_{33} {V^{U_R}_{32}}^*,
\label{eq:CH}
\end{eqnarray}
where $\alpha_s\simeq 0.1$ is the QCD coupling
evaluated at the mass scale of the sparticles in the loop,
$A_t$ is the soft SUSY breaking $A$ term for the top
 squark, $\mu$ denotes the two Higgs superfields mixing
parameter, $\tan \beta$ is the ratio of the two Higgs VEVs,
$m_{\tilde{g}}$ is the mass of the  gluino and
$V^{H}_{ij}$ is the mixing matrix in the charged-Higgs-squark
coupling $H^+ {\tilde{u_i}_R}^* \tilde{d_j}_L$.
The integral function $I_{H}$ is given by
\begin{eqnarray}
 I_{H} & = & \int_0^1 dz_1 \int_0^{1-z_1} dz_2 \frac{2}{
\frac{m_{\tilde{t}}^2}{m_{\tilde{g}}^2}z_1+
\frac{m_{\tilde{b}}^2}{m_{\tilde{g}}^2}z_2+
(1-z_1-z_2)},
\label{eq:IH}
\end{eqnarray}
which is equal to one at $m_{\tilde{t}}=m_{\tilde{b}}=m_{\tilde{g}}$
and varies slowly away from this degenerate point.
The contributions to $\Delta_S$ and $\Delta_P$
 from charged Higgs exchange are then given by
\begin{eqnarray}
\Delta_S & = & - \frac{\alpha_s}{3\pi} I_H \tan \beta
\frac{m_B}{(m_b-m_c)} \frac{m_B m_t}{m_H^2} \times
\frac{\mu + A_t\cot\beta}{m_{\tilde{g}}} \times
\frac{[V^{H}_{33} V^{D_L}_{33} {V^{U_R}_{32}}^*]}{V_{cb}} \, ,\\
\Delta_P & = &  \frac{\alpha_s}{3\pi} I_H \tan \beta
\frac{m_B}{(m_b+m_c)} \frac{m_B m_t}{m_H^2} \times
\frac{\mu + A_t\cot\beta}{m_{\tilde{g}}} \times
\frac{[V^{H}_{33} V^{D_L}_{33} {V^{U_R}_{32}}^*]}{V_{cb}}\, .
\label{eq:deltaSP}
\end{eqnarray}

To estimate the maximal $\tau$ polarization effects,
we assume $|V^{D_L}_{33}|=|V^{H}_{33}| \sim 1$,
$m_{H}=100\; \mbox{GeV}$ and
$\tan \beta=50$ \cite{tanbeta}.  With maximal squark mixings,
$|V^{U_R}_{32}|=1/\sqrt{2}$.
Setting $|\mu|=A_t=m_{\tilde{g}}$, $m_t=180\;\rm{GeV}$,
$m_b=4.5\;\mbox{GeV}$,
$m_c=1.5\;\mbox{GeV}$, $V_{cb}=0.04$, and $I_{H}=1$, we find
\begin{eqnarray}
|\Delta_S| & \le & 1.6 \\
 |\Delta_P|  & \le & 0.8.
\end{eqnarray}

 Averaging over the whole phase space gives, for
$B \rightarrow D \tau \overline{\nu}$,
\begin{eqnarray}
\left|\overline{P^{(D)}_{\tau}}\right| & = &
 0.22 \times \left|{\rm Im} \Delta_S\right| \le 0.35 \, ,
\end{eqnarray}
and, for $B \rightarrow D^* \tau \overline{\nu}$,
\begin{eqnarray}
\left|\overline{P^{(D^*)}_{\tau}}\right| & = &
 0.068 \times \left|{\rm Im} \Delta_P\right| \le 0.05 \, .
\end{eqnarray}
Both limits scale as $ \left( \case{100\text{ GeV}}{m_H} \right)^2
                \left(\frac{\tan \beta}{50}\right)
   \left(\frac{{\rm Im}[V^{H}_{33} V^{D_L}_{33} {V^{U_R}_{32}}^*]}
	{1/\sqrt{2}}\right)$.
In the absence of squark family mixing, the polarization effects are
suppressed by a factor of $\frac{m_tV^{U_R}_{32}}{m_bV_{cb}}\sim 10^3$.

\subsubsection{$W$ exchange and $D^*$ polarization}
\label{sec:W}

    As has been shown in section~\ref{sec:D*pol}, the $T$-odd polarization
 correlation of the $D^*$ in the decay $B\rightarrow  D^* \ell \overline{\nu}$
(with $\ell$$=$$e$, $\mu$)
is only sensitive to an effective right-handed (RH) quark current
interaction.
 With squark generational mixing,
an effective RH interaction can be induced
 at one loop by the $W$-boson exchange diagram
with left-right mass insertions
 in  both the top and bottom squark propagators.  This leads to a term
in the effective lagrangian given by \cite{wn2,wkn}
\begin{eqnarray}
{\cal L}_W & = &
 - \frac{4G_F}{\sqrt{2}} C_0
(\overline{c}_R \gamma^{\alpha} b_R)(\overline{\ell}_L
\gamma_{\alpha}\nu_L)
 + \rm{H.c.} \, , \label{eq:L1}
\end{eqnarray}
with
\begin{eqnarray}
C_0 & = & \frac{\alpha_s}{36 \pi} I_0
         \frac{m_tm_b(A_t-\mu\cot\beta)(A_b-\mu\tan\beta)}
              {m_{\tilde{g}}^4}
 V^{SKM}_{33}{V^{U_R}_{32}}^*V^{D_R}_{33}  ,
 \label{eq:C0}
\end{eqnarray}
where $A_b$ is the soft SUSY breaking $A$ term for the bottom squark,
$V^{SKM}_{ij}$ is the super CKM matrix associated with the
$W$-squark coupling $W^+ {\tilde{u_i}_L}^* \tilde{d_j}_L$, and
 the integral function $I_0$  is given by
\begin{eqnarray}
I_0 & = &  \int_0^1 dz_1 \int_0^{1-z_1} dz_2 \frac{24z_1z_2}{[
\frac{m_{\tilde{t}}^2}{m_{\tilde{g}}^2}z_1+
\frac{m_{\tilde{b}}^2}{m_{\tilde{g}}^2}z_2+
(1-z_1-z_2)]^2}.  \label{eq:I0}
\end{eqnarray}
Note that $I_0=1$ for
$\frac{m_{\tilde{t}}}{m_{\tilde{g}}}=
\frac{m_{\tilde{b}}}{m_{\tilde{g}}}=1$,
but it increases rapidly to $\sim 8$ as the squark-to-gluino mass
ratios decrease to $\frac{m_{\tilde{t}}}{m_{\tilde{g}}}=
\frac{m_{\tilde{b}}}{m_{\tilde{g}}}= \frac{1}{2}$.

  The $\Delta_R$ parameter of Eq.~(\ref{eq:deltaR})
is then given by
\begin{equation}
\Delta_R =
 - \frac{\alpha_s}{36 \pi} I_0
         \frac{m_tm_b(A_t-\mu\cot\beta)(A_b-\mu\tan\beta)}
              {m_{\tilde{g}}^4}
 \frac{V^{SKM}_{33}{V^{U_R}_{32}}^*V^{D_R}_{33}}{V_{cb}} .
\end{equation}
To estimate the maximal size of $\Delta_R$ from
the $W$-exchange diagram, we take
$I_0=5$, $\tan \beta=50$, $A_t=A_b=|\mu|=m_{\tilde{g}}=200 \;
\mbox{GeV}$, and  $|V^{D_R}_{33}|=|V^{SKM}_{33}|=1$.
With maximal squark mixing ($|V^{U_R}_{32}|=\frac{1}{\sqrt{2}}$),
 we have the upper limit
\begin{eqnarray}
|\Delta_R| & \le & 0.08 \; .
\end{eqnarray}
The averages of the two TOPO's related to
the $D^*$ polarization are given in Eqs.~(\ref{eq:TOPO1})
and (\ref{eq:TOPO2}).  Choosing the optimal orientations
of the polarization vector in the two cases and inserting
the above bound on $|\Delta_R|$ yields the following upper
limits
\begin{eqnarray}
 \left| \overline{P^{(1)}_{D^*}}\right| & \le &
     0.02, \\
\left| \overline{P^{(2)}_{D^*}}\right| & <& 0.016 .
\end{eqnarray}
These limits for the $D^*$ polarization scale as
$\left( \frac{200\text{ GeV}}{M_{\rm SUSY}} \right)^2
 \left( \frac{\tan \beta}{50} \right)
 \left( \frac{I_0}{5} \right)
 \left( \frac{{\rm Im}[V^{SKM}_{33}{V^{U_R}_{32}}^*V^{D_R}_{33}]}{1/\sqrt{2}}
\right)$, where $M_{\rm SUSY}$ is the SUSY breaking scale.
 In the absence of squark intergenerational mixing, the $D^*$ polarization
effect will be suppressed by a factor of
$\frac{m_tV^{U_R}_{32}}{m_cV_{cb}} \sim 10^{3}$.

\subsection{$R$-parity Violating Theories}
\label{sec:Rbreaking}

The requirement of gauge-invariance does not uniquely specify
the form of the superpotential in a generic supersymmetric model.
In addition to the terms which are usually present, one could also add
the following terms:
\equation
	\lambda_{ijk}L_iL_jE_k^c +\overline{\lambda}_{ijk}L_iQ_jD_k^c
	+\lambda_{ijk}^{\prime\prime}U_i^cD_j^cD_k^c
	+\mu_iL_iH^{\prime},
\label{eq:superpot}
\endequation
where the coefficients could in general be complex and
where $i$, $j$ and $k$ are generation indices.  Note that we have omitted the
implicit sum over $SU(2)_L$ and $SU(3)_C$ indices and that
$\lambda_{ijk}$$=$$-\lambda_{jik}$ and
$\lambda_{ijk}^{\prime\prime}$$=$$-\lambda_{ikj}^{\prime\prime}$.
Of the four types of terms listed above,
the last one may be
rotated away by a redefinition of the $L$ and $H$ fields~\cite{HS}.

  The above $\lambda$ and $\overline{\lambda}$ terms violate lepton
number whereas the $\lambda^{\prime\prime}$ term violates baryon number.
All three terms may be forbidden by imposing a discrete symmetry called
$R$-parity \cite{Rparity}.
Alternatively, one can use the experimental data to place constraints on
these $R$-parity breaking couplings.
The most stringent constraints are on the
$\overline{\lambda}\lambda^{\prime\prime}$ combinations and
come from the non-observation of proton decay \cite{SV}.
To satisfy the proton stability requirement, one can also invoke
a discrete $Z_3$ symmetry called baryon parity which
naturally allows for the lepton number violating terms while forbidding
the baryon number violating $\lambda^{\prime\prime}$ term \cite{Bparity}.
For this reason, we will simply set $\lambda^{\prime\prime}_{ijk}$$=$$0$
in our analysis.  The $R$-parity-violating interactions
in the lagrangian may then be written in the mass basis of the component
fields as
\begin{eqnarray}
	{\cal L}^{R\!\!\!\!/} & = & -2\lambda_{ijk}\left[
		\overline{(\nu^{i}_L)^c}e_{L}^j\widetilde{e}_{R}^{k*}
		+\overline{e^{k}_R}e_{L}^j\widetilde{\nu}_{L}^{i}
		+\overline{e^{k}_R}\nu_{L}^i\widetilde{e}_{L}^{j}\right]
	\nonumber \\
	& &   	- \lambda_{ijk}^{\prime} \left[(V_{KM})_{jl}\left(
		\overline{(\nu^{i}_L)^c}d_L^l\widetilde{d}_{R}^{k*}
		+\overline{d^{k}_R}d_{L}^l\widetilde{\nu}_{L}^{i}
		+\overline{d^{k}_R}\nu_{L}^i\widetilde{d}_{L}^{l}\right)
	\right.	\nonumber \\
	& &	\left. \;\;\;\;\; -\left(
		\overline{(e^{i}_L)^c}u_L^j\widetilde{d}_{R}^{k*}
		+\overline{d^{k}_R}u_{L}^j\widetilde{e}_{L}^{i}
		+\overline{d^{k}_R}e_{L}^i\widetilde{u}_{L}^{j}\right)
		\right] + {\rm H.c.}
\label{eq:RvL}
\end{eqnarray}
The $\overline{\lambda}$ and $\lambda^{\prime}$ parameters are related
by unitary rotations in generation space~\cite{AG}.  Note that while the above
parameterization is not unique (one could, for example, put $V_{KM}^{\dagger}$
in the ``up'' sector rather than $V_{KM}$ in the ``down'' sector)
the physics itself is parameterization-independent.

Integrating out the relevant supersymmetric particles of
Eq.~(\ref{eq:RvL})
gives rise to two types of contributions to the quark-level
transition $b\to c \ell \overline{\nu}$.  The first type of contribution
has the SM $V-A$ structure and cannot interfere with the SM $W$-exchange
diagram in order to give rise to observable $T$-violating effects.
The second type of contribution can induce scalar and pseudoscalar
effective interactions.  The relevant effective interaction
for the $\tau$ mode is given by
\begin{eqnarray}
	{\cal L}^{R\!\!\!\!/}_{\rm eff} & = &
		-\frac{1}{2}\frac{\lambda_{3j3}\lambda^{\prime *}_{j23}}
			{(m_{\widetilde{e}_{L}^{j}})^2}
		\overline{c}(1+\gamma^5)b\overline{\tau}(1-\gamma^5)\nu_{\tau}
+ {\rm H.c.} ,
\end{eqnarray}
where summation over $j=1,2$ is implied.
The resulting expressions for the corresponding $\Delta$
parameters are then
\begin{eqnarray}
	\Delta_S & = &
		-\frac{1}{2}\frac{\lambda_{3j3}\lambda^{\prime *}_{j23}}
			{(m_{\tilde{e}_{L}^{j}})^2}
			\left(\frac{\sqrt{2}}{G_F V_{cb}}\right)
			\frac{m_B^2}{(m_b-m_c)m_{\tau}}
, \\
	\Delta_P & = &
		-\frac{1}{2}\frac{\lambda_{3j3}\lambda^{\prime *}_{j23}}
			{(m_{\tilde{e}_{L}^{j}})^2}
			\left(\frac{\sqrt{2}}{G_F V_{cb}}\right)
			\frac{m_B^2}{(m_b+m_c)m_{\tau}}
.
\end{eqnarray}
Setting the slepton masses to 100 GeV
we obtain the following estimates
\begin{eqnarray}
	\Delta_S & \simeq & -8\times 10^2
                        \frac{\lambda_{3j3}
			\lambda^{\prime *}_{j23}}
			{(m_{\tilde{e}_{L}^{j}}/100\;{\rm GeV})^2}
,  \\
	\Delta_P & \simeq & -4\times 10^2
                        \frac{\lambda_{3j3}
			\lambda^{\prime *}_{j23}}
			{(m_{\tilde{e}_{L}^{j}}/100\;{\rm GeV})^2} \, .
\end{eqnarray}

    The tau polarization is subject to constraints from present
experimental data.
The rare decay $K^+ \to \pi^+ \nu \overline{\nu}$ gives the bound
$|\lambda^{\prime *}_{j23}| < 0.01$ \cite{AG},
whereas $|\lambda_{133}| < 0.001$ from bounds on the neutrino mass
\cite{GRT} and $|\lambda_{233}| < 0.03$ from leptonic tau decays
\cite{BGH}.  We have assumed in each case a mass of 100 GeV for the
sparticles.  We thus arrive at the following $90\%$ confidence level
upper bounds on the transverse $\tau$ polarizations,
\begin{eqnarray}
\left|\overline{P_{\tau}^{(D)}}\right| & < & 0.05 \; ,
\label{eq:RtauD}
\end{eqnarray}
\begin{eqnarray}
\left|\overline{P_{\tau}^{(D^*)}}\right| & < & 0.008 \; .
\label{eq:RtauD*}
\end{eqnarray}
In the limit of degenerate sparticle masses, these bounds
are independent of the sparticle mass scale.

We noted above that there are actually two types of $R$-parity
violating processes which could contribute to the quark-level
transition $b\to c \ell \overline{\nu}$.  The first of these was
ignored since it has the SM $V-A$ structure and thus cannot interfere
with the SM $W$-exchange diagram, while the second was seen to give
rise to an effective scalar-pseudoscalar interaction.  It is interesting
to note, however, that the SM-like term can {\em also} give rise to a $T$-odd
transverse $\tau$ polarization
if it interferes with the tree-level charged-Higgs diagram which
is generically present in supersymmetric models.  This effect is technically
of second order in the $\Delta$ parameters, yet it need not be small if we
take the current upper limit on $\tan \beta /m_H$, which is approximately
$0.5$ GeV$^{-1}$~\cite{tanbeta}.  In this limit the magnitude of the effect
could be comparable to the limits quoted in
Eqs.~(\ref{eq:RtauD}) and (\ref{eq:RtauD*}).
Note also that while $R$-parity violating interactions can give
rise to scalar and pseudoscalar interactions,
there is no tree-level induced  right-handed
current interaction which could
contribute to the $T$-odd $D^*$ polarization.

\subsection{Non-supersymmetric Models}

  In this subsection, we estimate the contributions to the TOPO's
in some non-SUSY models. We will consider in turn
the three-Higgs-doublet model (3HDM), leptoquark models and left-right
symmetric models (LRSM's).

\subsubsection{Multi-Higgs-doublet model}

  An effective scalar-pseudoscalar four-Fermi interaction can be induced
by tree-level charged Higgs exchange with $C\!P$-violating complex couplings.
To be specific, let us consider the three Higgs-doublet model
\cite{Weinberg,3HDMbounds}.
The charged Higgs couplings to the fermions are given by
\begin{eqnarray}
{\cal L} & = & (2\sqrt{2}G_F)^{1/2} \sum_{i=1}^{2}
         ( \alpha_i \overline{U_L}V_{KM}M_DD_R
          +\beta_i \overline{U_R}M_U V_{KM} D_L
          + \gamma_i \overline{\nu_L}M_E E_R) H^+_i + \rm{H.c.},
\end{eqnarray}
where $M_U$, $M_D$, and $M_E$ are the diagonal mass matrices for the
up-type quarks, down-type quarks and charged leptons, respectively.
The complex couplings $\alpha_i$, $\beta_i$, and $\gamma_i$ appear in the
unitary mixing matrix between
the mass eigenstates and gauge eigenstates of the charged Higgs boson.
They satisfy six constraints, three of which are
\begin{eqnarray}
\frac{\rm{Im}(\alpha_1 \beta_1^*)}{\rm{Im}(\alpha_2 \beta_2^*)}
=\frac{\rm{Im}(\alpha_1 \gamma_1^*)}{\rm{Im}(\alpha_2 \gamma_2^*)}
=\frac{\rm{Im}(\beta_1 \gamma_1^*)}{\rm{Im}(\beta_2 \gamma_2^*)}
& = & -1 \; .
\end{eqnarray}
It is clear from these relations that the $C\!P$-violating
effective scalar and pseudoscalar interactions will always
be proportional to $(1/m^2_{H^+_1}-1/m^2_{H^+_2})$.
 Assuming that $H^+_2$ is much heavier than $H^+_1$,
we find that the scalar and pseudoscalar $\Delta$ parameters
are given by
\begin{eqnarray}
\Delta_S & = & \frac{(\alpha_1 \gamma_1^* m_b + \beta_1 \gamma_1^* m_c)
                 m_B^2}{m^2_{H^+_1} (m_b - m_c)} \; ,\\
\Delta_P & = & \frac{(\alpha_1 \gamma_1^* m_b - \beta_1 \gamma_1^* m_c)
                 m_B^2}{m^2_{H^+_1} (m_b + m_c)} \; .
\end{eqnarray}

  Current data place a more stringent bound on $\rm{Im}(\beta_1 \gamma_1^*)$
than on $\rm{Im}(\alpha_1 \gamma_1^*)$ \cite{3HDMbounds}.
For $m_{H^+_1} < 440 \; \rm{GeV}$, the inclusive process
$B \to X \tau \overline{\nu}$ gives the strongest limit of
$|\rm{Im}(\alpha_1 \gamma_1^*)|/m^2_{H^+_1} < 0.2 \; \rm{GeV}^{-2}$
 at the $95\%$ C.L. \cite{grossman}.
This limit in turn constrains the $\Delta$'s by
$|\rm{Im}\Delta_S| < 8$ and $|\rm{Im}\Delta_P| < 4$. Therefore, the
$\tau$ transverse polarizations in $B \rightarrow D \tau \overline{\nu}$
and $B \rightarrow D^* \tau \overline{\nu}$ decays are given
by
\begin{eqnarray}
|\overline{P^{(D)}_{\tau}}| & \le & \sim 1 \\
|\overline{P^{(D^*)}_{\tau}}| & < & 0.3 \; ,
\end{eqnarray}
which is in agreement with a previous estimate~\cite{garisto}. Qualitatively
similar results have been found in the inclusive case~\cite{atwood,grossman}.

\subsubsection{Leptoquarks}

Both scalar and vector leptoquark models \cite{leptoquark} can give rise
to effective
scalar and pseudoscalar interactions for the semileptonic $B$ decays.
The calculation of the $\tau$ transverse polarization in these
models is similar to the analysis of the
muon transverse polarization in $K^+ \rightarrow \pi^0 \mu^+ \nu$
decay~ \cite{BG}.  Unlike in that case, however, the current
experimental data allow for a rather large $\tau$ polarization in $B$ decays.
The difference compared to $K_{\mu3}$ is that the bound on the
couplings for $B$ decay comes mainly from
$t \rightarrow c \tau^+ \tau^-$ and
is much weaker than that for the $K_{\mu3}$ decay, which comes from
$D \rightarrow \mu^+ \mu^-$.

Let us consider, as an example, the following
$SU(3)_C \times SU(2)_L \times U(1)_Y$ invariant leptoquark interaction,
\begin{eqnarray}
{\cal L} & = & ({\lambda}_{ij} \overline{Q}_{i} {e_R}_j
            + {\lambda}^{\prime}_{ij} \overline{u_R}_{i} L_j ) \phi
            + \rm{H.c.} \; ,
\end{eqnarray}
where $Q$ and $L$ denote the usual quark and lepton doublets, respectively,
$\phi$ is a color-triplet, weak-doublet scalar leptoquark,
and $i,j$ are the family indices.
An effective scalar-pseudoscalar four-Fermi interaction is then induced
by the exchange of the scalar leptoquark, giving\footnote{We neglect
for simplicity the effective tensor interaction which is also induced
by this exchange.},
\begin{eqnarray}
{\cal L}_{\rm{eff}} & = & - \frac{1}{2}
            \frac{\lambda^*_{33} \lambda^{\prime}_{23}}{m^2_{\phi}}
            (\overline{c_R} b_L) (\overline{\tau_R} \nu_{L \tau}) .
\end{eqnarray}
The resulting expressions for $\Delta_S$ and $\Delta_P$ are
given by
\begin{eqnarray}
\Delta_S & = &- 50 \times \lambda^*_{33} \lambda^{\prime}_{23}
             \times \left(\frac{200 \; \rm{GeV}}{m_{\phi}}\right)^2 \\
\Delta_P & = & - \frac{m_b-m_c}{m_b+m_c} \Delta_S \; ,
\end{eqnarray}
so that the transverse $\tau$ polarization in
$B \rightarrow D \tau \overline{\nu}$ and
$B \rightarrow D^* \tau \overline{\nu}$ decays
can be respectively of order unity and $0.2$  if we take
$|\rm{Im}(\lambda^*_{33} \lambda^{\prime}_{23})|\sim 0.1$.
Note that leptoquark exchange does not give rise to a right-handed
current at tree level.

\subsubsection{Left-right symmetric models}

 An effective right-handed quark current can be induced at tree level
 in left-right symmetric models (LRSM's) \cite{LRSM}.
We will concentrate on this effect and neglect the effective scalar
and pseudoscalar interactions by assuming that the charged Higgs decouple.
Consider the most general class of models with gauge group
$SU(2)_L \times SU(2)_R \times U(1)$.
The charged gauge boson mass eigenstates are related to the weak eigenstates
by the following two by two unitary matrix,
\begin{eqnarray}
\left( \begin{array}{c}
        W^+_L \\ W^+_R \end{array} \right) & = &
\left( \begin{array}{cc}
        \cos \zeta & -\sin \zeta \\
        e^{i\omega} \sin \zeta & e^{i\omega} \cos \zeta \end{array}
           \right)
\left( \begin{array}{c}
        W^+_1 \\ W^+_2 \end{array} \right) \; ,
\label{eq:wlwrmix}
\end{eqnarray}
where $\zeta$ is the $W_L - W_R$ mixing angle and $\omega$ is a
$C\!P$-violating phase.  The bounds on $m_{W_2}$ and $\zeta$ depend on the
relation between the CKM mixing matrix for the left-handed quarks,
$V^L=V_{KM}$, and the analogous
mixing matrix $V^R$ for the right-handed quarks.
In any case, $m_{W_2}$ is at least heavier than several hundred GeV
\cite{BBS,LS}, and we can safely neglect its effect for the purposes
of our estimate.

The presence of the off-diagonal term in the $W_L-W_R$ mixing matrix
means that the lighter mass eigenstate $W_1$ can induce an effective
right-handed current interaction of the
form $(\overline{c_R} \gamma_{\mu} b_R)
(\overline{\ell_L} \gamma^{\mu} \nu_L)$.
The resulting expression for $\Delta_R$ has the simple form
\begin{eqnarray}
\Delta_R & = & - e^{i\omega} \zeta \frac{g_R V^R_{cb}}{g_L V^L_{cb}} \; ,
\end{eqnarray}
where $g_L$ and $g_R$ are the gauge couplings for $SU(2)_L$ and $SU(2)_R$,
respectively. We will assume $g_L=g_R$ for our estimate.

  Stringent bounds on the $W_L - W_R$ mixing have been derived by
assuming manifest left-right symmetry ($V^L=V^R$) or
pseudo-manifest left-right symmetry ($V^R=K_1(V^L)^*K_2$, where $K_1$
and $K_2$ are diagonal phase matrices).  Thus,
for example, $|\zeta| < 4\%$ from $\mu$ decay experiments \cite{mudecay},
$|\zeta| < 4 \times 10^{-3}$ from the analysis of $K \rightarrow 2\pi$
and $K \rightarrow 3\pi$ decays (subject to some theoretical hadronic
uncertainties) \cite{DH},
and $|\zeta| < 5 \times 10^{-3}$ from semileptonic $d$ and $s$ decays
\cite{wolf}.  The upper bound on $|\rm{Im} \Delta_R|$ is then
in the range
\begin{eqnarray}
|\rm{Im} \Delta_R| & \le & |\zeta|
< (0.004 \sim 0.04) \; ,
\end{eqnarray}
and the $D^*$ polarization in these scenarios is
smaller than $10^{-3} \sim 10^{-2}$.

  If one does not impose manifest or pseudo-manifest left-right symmetry,
the constraints on $\zeta$ tend to become less stringent. Thus,
for example, it is possible to have $|V^R_{cb}|=1$ and $|\zeta| \le 0.013$
at the $90\%$ C.L. \cite{LS}.
The induced right-handed current can be significantly
enhanced in this case, since
\begin{eqnarray}
|\rm{Im} \Delta_R| & \le & 25 \times |\zeta| \le 0.32
\end{eqnarray}
and the $T$-odd $D^*$ polarization in the
$B \rightarrow D^* \ell \overline{\nu}$ ($\ell=e,\mu$) decays
could be as large as $8\%$.

\section{Discussion and Conclusions}
\label{sec:conclusion}

In this paper we have examined several of the $T$-odd
polarization observables in the exclusive
semileptonic decays $B \rightarrow D^{(*)} \ell \overline{\nu}$.
We have provided a model-independent analysis of these
observables, concentrating on the $\tau$ transverse polarization
in $B \rightarrow D^{(*)} \tau \overline{\nu}$ and on the $T$-odd
$D^*$ polarization in the decays $B \rightarrow D^{*} \ell \overline{\nu}$,
with $\ell=e,\;\mu$.  These observables provide an attractive place
in which to look for effects coming from new physics.  As is known, they
receive negligible contributions from standard model sources.  Furthermore,
they are quite clean theoretically, depending only on a small number of
$q^2$-dependent form factors which are in principle measurable or calculable
on the lattice or within the context of Heavy Quark Effective Theory.
We have also noted that the three
types of observables under consideration are sensitive separately
to three different types of quark-level effective interactions:
the $\tau$ polarization in the decay to the $D$ ($D^*$) probes
effective scalar (pseudoscalar) interactions, and the $T$-odd
$D^*$ polarization depends only on effective right-handed current
interactions.  This observation is independent of the functional forms of
the form factors.  A final general remark concerning these observables is that
the branching ratios for these decays should be quite accessible at
the planned $B$-factories.  Using the leading order results of HQET and taking
$\xi(w)=1.0 -0.75 \times (w-1)$ (as we have in our numerical work),
we find that
\equation
B(B\!\to\! D \tau \overline{\nu})\!:\!
B(B\!\to\! D^* \tau \overline{\nu})\!:\!
B(B\!\to\! D \ell \overline{\nu})\!:\!
B(B\!\to\! D^* \ell \overline{\nu})
\sim \frac{1}{10}:\frac{1}{4}:\frac{1}{3}:1 \; ,
\endequation
with $\ell=e$ or $\mu$.  While these
ratios should be taken as being only approximate, they
do indicate that one can expect branching
ratios for the first two decays (which are currently unmeasured)
to be of order one percent.  They
also show that, all else being equal, the
experimental sensitivity to a $T$-violating effective right-handed current
interaction is much greater than that to a scalar or
pseudoscalar interaction.  This is particularly true if one combines
the measurements in the electron and muon modes.

In this work we have not included the effects of possible
tensor interactions.  In all of the models which we have considered --
with the possible exception of the leptoquark models -- such effects are
either not present or are quite small.  It is worth noting, however,
that a model-independent analysis of tensor effects may also be
performed along the same lines as followed here \cite{kw}.  It is also
straightforward to derive the
tensor form factors for both the $B \rightarrow D$ and
$B\rightarrow D^*$ transitions in HQET.

It is interesting to compare the sensitivity of the tau transverse
polarization in $B_{\tau3}$ to that of the muon in $K^+_{\mu3}$.
A priori one expects the polarization effect to be larger for $B_{\tau3}$
than for $K^+_{\mu3}$ due to the larger quark and lepton
masses in the $B$ case.  The lepton polarization in these two cases
may generically be written as
$P_\ell \sim \sigma_\ell \times \rm{Im} \Delta_S^\ell$, where the
kinematical polarization function $\sigma_\ell$ contains a helicity
suppression factor,
$\sigma_{\ell} \propto m_{\ell}/m_M$ ($m_M$ is the mass of the decaying
meson), and where $\Delta_S^\ell$ is a model-dependent
parameter which measures the
strength of the effective scalar interaction.
The relative sizes of $\Delta_S^\tau$ and $\Delta_S^\mu$ are model-dependent,
so let us consider the 3HDM as an example.  In this case the ratio
$\Delta_S^\tau/\Delta_S^\mu$ is enhanced roughly by the factor $m_B^2/m_K^2$.
Thus, up to numerical factors of order unity,
the transverse lepton polarization is enhanced by
$P_{\tau}/P_{\mu} \sim m_Bm_{\tau}/m_Km_{\mu}\sim 10^2$.
Similar qualitative analyses
can be performed for the other models which we have considered.
The rather large enhancement which one generically finds implies
that in order to reach a given sensitivity to new physics, one requires
far fewer $B$ decays than $K$ decays.  The $B$
system, as we have noted above, has the added advantage
that there are several semileptonic $B$
decay channels which have no analogue in the $K$ system and which
may in principle be used to identify separately the various
possible sources of $T$ violation.

Although we have considered here only the decays
$B \rightarrow D^{(*)} \ell \overline{\nu}$, our
results may also be applied to the related decays
$B \rightarrow \pi (\rho, \omega) \ell \overline{\nu}$.
The results of HQET are not applicable to these decays, so that
the form factors need
to be obtained using phenomenological models and/or experimental data.
It is expected, however, that the $T$-odd
polarization effects in these modes could
be just as large as for the $b\rightarrow c$ transitions.  The usefulness
of these decays as probes for $T$-odd signals of new physics may be
limited, however, since their branching ratios are expected to be
smaller by one to two orders of magnitude.

In conclusion, we have presented a general analysis of several
$T$-odd polarization observables in
the semileptonic $B$ decays to $D$ and $D^*$
mesons.  We have given numerical estimates of theses observables
in both supersymmetric $R$-parity conserving and $R$-parity breaking
models as well as in some non-supersymmetric
extensions of the SM, namely the three-Higgs-doublet model,
leptoquark models, and left-right symmetric models.  The results of these
model estimates have been summarized in Table~\ref{tab:models}.
It is encouraging that the polarization effects in
many of these models can be in the range of a few percent to several
tens of percent and could thus be accessible to the
planned $B$ factories.

\acknowledgments
{We would like to thank Drs. M. Atiya, D. Coward, S. Dawson, M. Diwan,
R. Garisto, F. Goldhaber, Y. Grossman, Y. Kuno, K. Lau, W. Marciano, F. Paige,
M. Pospelov and A. Soni for useful conversations. This work is partially
supported by the Natural Sciences and Engineering Research Council of Canada.
K.K. is also grateful to the High Energy Theory Group at Brookhaven National
Laboratory for support provided under contract number DE-AC02-76CH00016 with
the U.S. Department of Energy.}

\newpage

\appendix

\section{}
\label{sec:appenda}

In this appendix we define the kinematical functions $g_i(x,y)$ and $f_i(x,y)$
which arise in the definitions of $\rho_D(x,y)$ and $\rho_{D^*}(x,y)$.
They are given, for lepton $\ell$, by:
\begin{mathletters}
\begin{eqnarray}
g_1(x,y) & = & (3-x-2y +r_{\ell} - r_D)(x+2y-1-r_{\ell} -r_D)
      \nonumber \\ &&
            -(1+x+r_D)(1-x+r_D-r_{\ell}) \\
g_2(x,y) & = & r_{\ell}(3-x-2y-r_D + r_{\ell}) \\
g_3(x) & = & r_{\ell}(1-x+r_D - r_{\ell}) \,
\end{eqnarray}
\end{mathletters}
and
\begin{mathletters}
\begin{eqnarray}
f_1(x,y) & = & (1-x+r_{D^*} - r_{\ell})
             + \frac{1}{r_{D^*}} (x+y-1-r_{D^*} - r_{\ell})
		(1-y + r_{\ell} -r_{D^*})
 \\
f_2(x,y) & = &  [ (x+2y-1-r_{D^*} -r_{\ell})(3-x-2y-r_{D^*}+r_{\ell})
 \nonumber \\
 &&                -(1-x+r_{D^*} - r_{\ell})(1+x+r_{D^*})]
               \left( \frac{x^2}{4r_{D^*}}-1 \right)
 \\
f_3(x) & = & r_{\ell} (1-x+r_{D^*} - r_{\ell})
          \left( \frac{x^2}{4r_{D^*}}-1 \right)
\\
f_4(x,y) & = &   2xy(1-y + r_{\ell} -r_{D^*})
              + 2 x(2-x-y)(x+y-1-r_{D^*} - r_{\ell})
 \nonumber \\
  &&            - 4(1-y + r_{\ell} -r_{D^*})(x+y-1-r_{D^*} - r_{\ell})
              - 4r_{D^*} y(2-x-y)
 \\
f_{5}(x,y) & = & \frac{1}{r_{D^*}} x(1-y)(x+y-1)
                 - \frac{r_{\ell}}{2r_{D^*}}x(3-2x-3y -r_{D^*} + r_{\ell})
 \nonumber \\
 &&                + 2(1-y)(1-x-y) -x + 2r_{D^*} -r_{\ell}(x+y)
 \\
f_{6}(x,y) & = & \frac{r_{\ell}}{2r_{D^*}}[x(1-y + r_{\ell} -r_{D^*})
                                       -2r_{D^*}(2-x-y)]
\\
f_{7}(x,y) & = & r_{\ell}(3-x-2y-r_{D^*} +r_{\ell})
                  \left( \frac{x^2}{4r_{D^*}}-1 \right)
\\
f_{8}(x,y) & = & 2y(1-y + r_{\ell} -r_{D^*})
                  -2(2-x-y)(x+y-1-r_{D^*} - r_{\ell}) .
\end{eqnarray}
\end{mathletters}

\section{Four-body Final States}
\label{sec:appendb}

In this appendix we demonstrate how the two $T$-odd
$D^*$ polarization observables defined in the text
(see Eqs.~(\ref{eq:poldstarl}), (\ref{eq:TOPO1}) and (\ref{eq:TOPO2}))
may be related to $T$-odd momentum correlations in the four-body
final state of the decay $B$$\to$$D^*(D\pi)\ell\overline{\nu}$.
The two observables have different structures in terms of the
$D^*$ polarization vector and
may be separately extracted by employing suitable integration
prescriptions in the integration over the momentum of the final
state pion.
We will examine two different types of prescriptions and calculate
the statistical error in each case.
A previous analysis of $T$-odd asymmetries in the four-body final
state may be found in Refs.~\cite{GV,KSW}, where it was noted that the
final state interaction effects on the $T$-odd observables are
probably negligible.  One could similarly study the $T$-odd momentum
correlations in the channel
$B$$\to$$D^*(D\gamma)\ell\overline{\nu}$, but this channel will
not be examined here.

Let us then calculate the differential partial width for
$B$$\to$$D^*(D\pi)\ell\overline{\nu}$.
The Feynman rule for the effective
$D^{*\mu}$-$\pi$-$D$ vertex is simply given by
$fp_{\pi}^{\mu}$~\cite{ddpivertex}, where the constant $f$
may be inferred from the partial
width of the decay $D^*\rightarrow\pi D$.  This width is given by
\equation
	\Gamma(D^*\rightarrow \pi D) = \left(\frac{1}{3}\right)
			\frac{|f|^2(p_{\pi})^3}
			{8\pi m_{D^*}^2},
\endequation
where
\equation
	p_{\pi} = \frac{1}{2m_{D^*}}\lambda^{1/2}(m_{D^*}^2,m_{\pi}^2,
		m_D^2)
\endequation
denotes the magnitude of the pion momentum in the $D^*$ rest frame and
$\lambda(x,y,z)$$=$$x^2+y^2+z^2-2xy-2xz-2yz$.  In order to
calculate the decay rate for
$B$$\to$$D^*(D\pi)\ell\overline{\nu}$, we need to sum over the intermediate
states of the $D^*$, which may be done either by using a Breit-Wigner
propagator for the $D^*$ or by employing a density matrix approach.
The resulting expression for the partial differential width
in the $B$ rest frame is given by
\equation
	\left.\frac{d^2\Gamma(B\to D^*(D\pi)\ell\overline{\nu})}{dxdy}\right|_S
		 =  \frac{3m_BG_F^2|V_{cb}|^2}{512\pi^4(p_{\pi}^*)^2}
		\left(\int_S d\Omega_{\pi}^*
		\left|\widetilde{{\cal M}}\right|^2 \right)
		\times{\rm BR}(D^*\rightarrow\pi D),
	\label{eq:widthdef}
\endequation
where
\equation
	\widetilde{{\cal M}} =
		{\cal M}^{\rho\alpha}
		\left( p_{\pi\rho}-p_{D^*\rho}\frac{p_{\pi}\cdot p_{D^*}}
			{m_{D^*}^2}\right)
		\overline{u}_L(p_{\ell})\gamma_{\alpha} v_L(p_{\nu}),
\endequation
and where ${\cal M}^{\rho\alpha}$ has been defined above in
Eq.~(\ref{eq:mrhomu}).
The angular integral in Eq.~(\ref{eq:widthdef}) is to be performed
{\it in the rest frame of the decaying} $D^*$ using some prescription
``$S$.'' This prescription may be designed such that it picks out the
$T$-odd contributions.

The angles in the $D^*$ rest frame may be defined as follows
\begin{eqnarray}
	\vec{p}_B & = & \left|\vec{p}_B\right|\left(0,0,-1\right) \; , \\
	\vec{p}_\ell & = & \left|\vec{p}_\ell\right|\left(\sin\theta_\ell,
		0,\cos\theta_\ell\right) \; , \\
	\vec{p}_\pi & = & \left|\vec{p}_\pi\right|\left(
		\sin\theta_\pi \cos\phi_\pi , \sin\theta_\pi \sin\phi_\pi ,
		\cos\theta_\pi \right) ,
\end{eqnarray}
where $\vec{p}_B$, $\vec{p}_\ell$ and
$\vec{p}_\pi$ are the momenta {\em in the rest frame of the} $D^*$.
There are then in principle three $T$-odd structures which one may construct
in terms of the pion momentum.  These are
\begin{eqnarray}
	\vec{p}_{\pi}\cdot \left(\vec{p}_B\times\vec{p}_{\ell}\right)
		& \sim & \sin\theta_{\pi} \sin\phi_{\pi} ,\\
	\left(\vec{p}_{\pi}\cdot \vec{p}_{\ell}\right)
		\vec{p}_{\pi}\cdot \left(\vec{p}_B\times\vec{p}_{\ell}\right)
		& \sim & \sin\theta_{\pi}\cos\theta_{\pi} \sin\phi_{\pi},\;\;
		\sin^2\theta_{\pi}\sin\phi_{\pi}\cos\phi_{\pi} ,\\
	\left(\vec{p}_{\pi}\cdot \vec{p}_B\right)
		\vec{p}_{\pi}\cdot \left(\vec{p}_B\times\vec{p}_{\ell}\right)
		& \sim & \sin\theta_{\pi}\cos\theta_{\pi} \sin\phi_{\pi}.
\end{eqnarray}
Only the latter two structures are present in the partial width since, in the
$D^*$ rest frame,
\equation
	\left( p_{\pi\rho}-p_{D^*\rho}\frac{p_{\pi}\cdot p_{D^*}}
			{m_{D^*}^2}\right) \longrightarrow
	g_{\rho i} p_{\pi}^i,
\endequation
so that all terms in the squared amplitude are bilinear in the pion momentum.
The observable $T$-odd functional forms are then given by
\begin{eqnarray}
	T_1(\theta_{\pi},\phi_{\pi}) & = &
		\sin^2\theta_{\pi}\sin\phi_{\pi}\cos\phi_{\pi}, \\
	T_2(\theta_{\pi},\phi_{\pi}) & = &
		\sin\theta_{\pi}\cos\theta_{\pi} \sin\phi_{\pi} .
\end{eqnarray}

There are several integration prescriptions which may be used
to extract the terms in the width which are proportional to
$T_1$ and $T_2$.  In general these reduce to weighting the
differential width by some function $f(\theta_{\pi},\phi_{\pi})$ in such
a way that only the desired piece survives the angular integration.
We shall examine two such prescriptions in this appendix.  The first
approach (prescription ``A'') is closely related to
that used in Ref.~\cite{KSW} and amounts
to weighting the angular integral by $\pm 1$, depending on the angle.
In the second approach (prescription ``B''), the integrand is weighted
by the functional form itself, which also has
the effect of eliminating all but the
desired piece.  As we shall show, prescription ``B'' is
statistically more efficient than prescription ``A.''

Let us first consider prescription ``A''.  In this case
the integrand is weighted by $\pm 1$ as
a function of the angle.  Two different such prescriptions may be used
to separately pick out the terms proportional to $T_1$ and $T_2$, while
eliminating all other terms.
It is straightforward to verify that the following two prescriptions
do the job:
\begin{eqnarray}
	T_1(\theta_{\pi},\phi_{\pi})&:&\;\;
		\int_{A_1} d\Omega_{\pi}^*	\equiv
		\int_0^{\pi}\sin\theta_{\pi}d\theta_{\pi}
		\left(\int_0^{\pi/2} -\int_{\pi/2}^{\pi}
          +\int_{\pi}^{3\pi/2} -\int_{3\pi/2}^{2\pi}\right)d\phi_{\pi}, \\
	T_2(\theta_{\pi},\phi_{\pi})&:&\;\;
		\int_{A_2} d\Omega_{\pi}^*	\equiv
		\left(\int_0^{\pi/2} - \int_{\pi/2}^{\pi}\right)
		\sin\theta_{\pi}d\theta_{\pi}
		\left(\int_0^{\pi} -\int_{\pi}^{2\pi}\right)d\phi_{\pi} .
\end{eqnarray}
We may then define the following normalized asymmetries
\begin{eqnarray}
		{\cal A}_{A_1}(x,y) & \equiv &
	\left(\frac{d^2\Gamma_{A_1}^{\rm 4-bdy}}{dxdy}
			\right) \times
	\left(\frac{d^2\Gamma^{\rm 4-bdy}}{dxdy}\right)^{-1}
	\label{as1a} \\
	& = & -\frac{1}{\pi}\lambda_1(x,y)
		{\rm Im}\left(F^{\prime}_{A0}F_V^{\prime *}\right)
	\label{as1}
\end{eqnarray}
and
\begin{eqnarray}
	{\cal A}_{A_2}(x,y) & \equiv &
	\left(\frac{d^2\Gamma_{A_2}^{\rm 4-bdy}}{dxdy}
			\right) \times
	\left(\frac{d^2\Gamma^{\rm 4-bdy}}{dxdy}\right)^{-1}
	\label{as2a} \\
	& = & \frac{1}{\pi} \lambda_2(x,y)\left[
		{\rm Im}(F^{\prime}_{A0}F_{A+}^{\prime *})
		+ {\rm Im}(F_{A+}^{\prime}F_{V}^{\prime *})
		(x+2y-2-r_{\ell}) \right. \nonumber \\
		& & \left. \;\; +{\rm Im}
			(F_{A-}^{\prime}F_{V}^{\prime *})r_{\ell}
		+{\rm Im}(F_{A0}^{\prime}F_{V}^{\prime *})
			d_{\ell}(x,y)\right] \; .
	\label{as2}
\end{eqnarray}
The above two asymmetries are proportional to the two terms in the
expression given for the polarization of the $D^*$ in
Eq.~(\ref{eq:poldstarl}), that is,
\equation
	P^{(\ell)}_{D^*}(x,y) = (\vec{\epsilon}\cdot\vec{n}_2) \pi \left[
		(\vec{\epsilon}\cdot\vec{n}_1) {\cal A}_{A_1}(x,y)
		+ (\vec{\epsilon}\cdot\vec{n}_3) {\cal A}_{A_2}(x,y)
				\right] \, .
\endequation
We have thus confirmed our assertion that the two polarization structures
in Eq.~(\ref{eq:poldstarl}) may be measured separately by following the
decay of the $D^*$ and studying the $T$-odd momentum correlations
in the resulting four-body final state.

We now turn to prescription ``B''.  In this case the differential width
is weighted by the functional form itself in the angular integration.
One may easily verify that weighting the width by $T_i$
picks out the term proportional to
$T_i$ and eliminates all other terms.  Prescription ``B'' is then defined
by:
\begin{eqnarray}
	T_1(\theta_{\pi},\phi_{\pi})&:&\;\;
		\int_{B_1} d\Omega_{\pi}^* \equiv
		\int d\Omega_{\pi}^* \left(\frac{10}{\pi}\right)
		T_1(\theta_{\pi},\phi_{\pi}) \; ,\\
	T_2(\theta_{\pi},\phi_{\pi})&:&\;\;
		\int_{B_2} d\Omega_{\pi}^* \equiv
		\int d\Omega_{\pi}^* \left(\frac{10}{\pi}\right)
		T_2(\theta_{\pi},\phi_{\pi}) \; .
\end{eqnarray}
The normalizing factor of $10/\pi$ has been
included so that the resulting asymmetries (defined in analogy with
Eqs.~(\ref{as1a}) and (\ref{as2a})) have the same numerical value using either
method; that is, ${\cal A}_{B_i}(x,y)$$=$${\cal A}_{A_i}(x,y)$.

In order to compare prescriptions ``A'' and ``B'', it is useful to calculate
the statistical uncertainties which would be expected in a measurement of
the two asymmetries, ${\cal A}_{A_i}$ and ${\cal A}_{B_i}$,
given some number of events $N$.  In particular, we will calculate
the uncertainties of the averaged quantities $\overline{{\cal A}_{A_i}}$
and $\overline{{\cal A}_{B_i}}$, in which the averages over $x$ and $y$
are performed as prescribed in Eqs.~(\ref{eq:TOPO1}) and
(\ref{eq:TOPO2}), for $i=1$ and $2$, respectively.  The numerical
calculations will be carried out for the electron and muon channels,
since these are the modes which we have concentrated on in the text.

We first define the expectation value of some
operator ${\cal O}$ as follows:
\equation
	\langle {\cal O} \rangle \equiv
		\frac{\int dx dy \int d\Omega_{\pi}^*
		\left|\widetilde{{\cal M}}\right|^2 {\cal O}}
		{\int dx dy \int d\Omega_{\pi}^*
		\left|\widetilde{{\cal M}}\right|^2} .
\endequation
This expectation value corresponds to a ``measurement'' of the
operator ${\cal O}$ in the probability distribution defined by
$\left|\widetilde{{\cal M}}\right|^2$.
The statistical error for this observable, given $N$ events, is then
\equation
	\sigma_{\cal O} = \frac{\sqrt{\langle {\cal O}^2 \rangle
		-\langle {\cal O} \rangle^2}}{\sqrt{N}}.
	\label{eq:staterr}
\endequation
The four averaged asymmetries may be expressed in terms of this
compact notation by writing
\equation
	\overline{{\cal A}_{A_i}}  =  \langle {\cal O}_{A_i} \rangle \; ,
	\;\;\;\;\;\;
	\overline{{\cal A}_{B_i}}  =  \langle {\cal O}_{B_i} \rangle \; ,
\endequation
where the four operators are given by
\equation
	{\cal O}_{A_i}  =  \pm 1 \; ,
	\;\;\;\;\;\;
	{\cal O}_{B_i}  =  \pm\left(\frac{10}{\pi}\right)
		T_i(\theta_{\pi},\phi_{\pi}) \; .
\endequation
The appropriate sign to choose in the above expressions depends in general
on $\theta_{\pi}$, $\phi_{\pi}$ and $y$.

It is now straightforward to calculate the statistical uncertainties
associated with the averaged asymmetries in the prescriptions ``A''
and ``B''.  In order to evaluate these numerically, we may safely
neglect the term $\langle {\cal O}\rangle^2$ in Eq.~(\ref{eq:staterr}),
since it is the square of the averaged asymmetry and is typically quite small
compared to $\langle {\cal O}^2\rangle$, which is of order unity.
Taking $\xi(w)=1.0 -0.75 \times (w-1)$ and setting the $\Delta$'s to
zero in $\left|\widetilde{{\cal M}}\right|^2$, we find
\begin{eqnarray}
	\sigma_{A_1} & \simeq & \frac{\sqrt{\langle {\cal O}_{A_1}^2 \rangle}}
		{\sqrt{N}} = \frac{1}{\sqrt{N}} \; ,\\
	\sigma_{A_2} & \simeq & \frac{\sqrt{\langle {\cal O}_{A_2}^2 \rangle}}
		{\sqrt{N}} = \frac{1}{\sqrt{N}} \; ,\\
	\sigma_{B_1} & \simeq & \frac{\sqrt{\langle {\cal O}_{B_1}^2 \rangle}}
		{\sqrt{N}} = \frac{0.75}{\sqrt{N}} \; ,\\
	\sigma_{B_2} & \simeq & \frac{\sqrt{\langle  {\cal O}_{B_2}^2\rangle}}
		{\sqrt{N}} = \frac{0.89}{\sqrt{N}} .
\end{eqnarray}
We could, alternatively, calculate the number of events required to
achieve a given statistical uncertainty.  In this case, the ratio of the
number of events required in prescriptions ``B'' and ``A'' is given
by
\begin{eqnarray}
	\frac{N_{B_1}}{N_{A_1}} & \simeq & 0.57 \; , \\
	\frac{N_{B_2}}{N_{A_2}} & \simeq & 0.79 \; .
\end{eqnarray}
Thus prescription ``B'' is more efficient than prescription
``A'', as we have asserted.

\newpage
\begin{table}
\caption{$T$-odd polarization observables (TOPO's) for exclusive
semileptonic $B$ decays
in terms of effective  scalar ($\Delta_S$), pseudoscalar ($\Delta_P$),
right-handed quark current ($\Delta_R$), and
left-handed quark current ($\Delta_L$) four-Fermi interactions.
$P_{\tau}^{(D)}$ and
$P_{\tau}^{(D^*)}$ denote the transverse $\tau$ polarization in the
$B \rightarrow D \tau \overline{\nu}$ and
$B \rightarrow D^* \tau \overline{\nu}$ decays respectively;
$P_{D^*}^{(\ell)}$ ($\ell$$=$$e$,$\mu$,$\tau$) denotes the $T$-odd $D^*$
polarization observable in the
$B \rightarrow D^* \ell \overline{\nu}$ decay.}
\label{tab:general}
\begin{tabular}{ccccc}
   & $\rm{Im}\Delta_S$ & $\rm{Im}\Delta_P$ & $\rm{Im}\Delta_R$  &
     $~~~~\rm{Im}\Delta_L~~~~$ \\ \hline
$P_{\tau}^{(D)}$  & $\surd$ &  0 & 0 & 0  \\
$P_{\tau}^{(D^*)}$ & 0 &  $\surd$  & 0 & 0 \\
$P_{D^*}^{(e,\mu)}$ & 0 & 0 & $\surd$ & 0 \\
$P_{D^*}^{(\tau)}$     & 0  & $\surd$  & $\surd$ & 0
\end{tabular}
\end{table}

\vskip 0.3in

\begin{table}
\caption{Contributions to the effective four-Fermi interactions
and to the various TOPO's from SUSY with squark intergenerational mixing,
SUSY with $R$-parity violation, the three Higgs-doublet model (3HDM),
leptoquark models, and left-right symmetric models (LRSM's).
We have neglected the effects due to charged Higgs bosons in LRSM's, assuming
that the Higgs bosons are sufficiently heavy  to decouple.
The numbers in the table
are the maximal polarization effects and are meant mainly for the purpose
of illustration.  Their actual sizes in particular models will
depend on the details of the models. }
\label{tab:models}
\begin{tabular}{cccccc}
     & squark mixing  & $R\!\!\!\!/\!\;$ SUSY & 3HDM  &
	Leptoquarks  & LRSM  \\ \hline
$\Delta_S$  & $\surd$  & $\surd$  &  $\surd$  &  $\surd$  & 0  \\
$\Delta_P$  & $\surd$  & $\surd$  &  $\surd$  &  $\surd$  & 0    \\
$\Delta_R$  & $\surd$  & 0  & 0   &  0  & $\surd$   \\ \hline
$|P_{\tau}^{(D)}|$  & 0.35 &  0.05 & $\sim 1$ &  $\sim 1$ & 0 \\
$|P_{\tau}^{(D^*)}|$  & 0.05 & 0.008 &  0.3 &
   0.2 & 0     \\
$|P_{D^*}^{(1)}|$    & 0.02  &  0 & 0 & 0 & 0.08  \\
$|P_{D^*}^{(2)}|$    & 0.016  & 0 &  0 & 0 & 0.06
\end{tabular}
\end{table}

\newpage

\newpage

\newpage

\end{document}